\documentclass[%
aip,
amsmath,amssymb,
reprint, onecolumn 
]{revtex4-1}
\usepackage{graphicx}
\usepackage{dcolumn}
\usepackage{bm}
\usepackage[utf8]{inputenc}
\usepackage[T1]{fontenc}
\usepackage{mathptmx}
\usepackage{etoolbox}
\usepackage{amscd,amsthm}
\usepackage{enumerate}

\usepackage[usenames,dvipsnames]{color}

\newcommand{\ind}{{\mathrm{Ind}}}

\newcommand{\BZ}{{\mathrm{\Omega}}}
\newcommand{\IBZ}{{\mathrm{\Omega_0}}}

\newcommand{\tr}{{\mathrm{Tr}}}

\newcommand{\bbC}{{\mathbb{C}}}

\newcommand{\bbR}{{\mathbb{R}}}
\newcommand{\bbN}{{\mathbb{N}}}
\newcommand{\bbZ}{{\mathbb{Z}}}
\newcommand{\bfa}{{\mathbf{a}}}
\newcommand{\bfb}{{\mathbf{b}}}

\newcommand{\bfk}{{\mathbf{k}}}

\newcommand{\bfm}{{\mathbf{m}}}
\newcommand{\bfn}{{\mathbf{n}}}

\newcommand{\bfe}{{\mathbf{e}}}

\newcommand{\tildef}{{\tilde{f}}}
\newcommand{\bfkn}{\bfk_\bfn}
\newcommand{\bfkm}{\bfk_\bfm}

\newcommand{\bft}{{\mathbf{t}}}
\newcommand{\bftm}{\bft_\bfm}
\newcommand{\bftn}{\bft_\bfn}

\newcommand{\bfv}{{\mathbf{v}}}
\newcommand{\bfx}{{\mathbf{x}}}
\newcommand{\bfy}{{\mathbf{y}}}

\newcommand{\calC}{{\mathcal{C}}}

\newcommand{\calF}{{\mathcal{F}}}
\newcommand{\hatG}{{\widehat{G}}}

\newcommand{\calH}{{\mathcal{H}}}

\newcommand{\calO}{{\mathcal{O}}}

\newcommand{\un}{{\mathrm{Id}}}

\newcommand{\id}{{\mathrm{Id}}}

\newcommand{\cgop}[2]{\{#1,#2\}}

\newcommand{\unit}{{e}}

\newcommand{\ppmatrix}[4]{\left( \begin{array}{cc}
#1 & #2 \\
#3 & #4
\end{array}\right)}

\newcommand{\bfq}{\mathbf{q}}
\newcommand{\bfr}{\mathbf{r}}

\newtheorem{prop}{Proposition}

\newtheorem{dfn}[prop]{Definition}

\makeatletter
\def\@email#1#2{%
\endgroup
\patchcmd{\titleblock@produce}
{\frontmatter@RRAPformat}
{\frontmatter@RRAPformat{\produce@RRAP{*#1\href{mailto:#2}{#2}}}\frontmatter@RRAPformat}
{}{}
}%
\makeatother
\begin{document}
\preprint{AIP/123-QED}

\title[Sample title]{Sample Title:\\with Forced Linebreak}
\title[Representations of the symmetry groups of infinite crystals]{Representations of the symmetry groups of infinite crystals}
\author{Bachir Bekka}
\affiliation{Universit\'e de Rennes, CNRS, Institut de recherche math\'ematique de Rennes,
IRMAR, UMR 6625, Campus Beaulieu, 35042 Rennes cedex, France.}
\author{Christian Brouder}
\affiliation{%
Sorbonne Universit\'e, CNRS, Mus\'eum National d'Histoire Naturelle, 
Institut de min\'eralogie, de physique des mat\'eriaux et de cosmochimie, IMPMC, 75005 Paris, France.
}%

\date{\today}

\begin{abstract}
We investigate the representations of the symmetry groups of infinite crystals.
Crystal symmetries are usually described as the finite symmetry
group of a finite crystal with periodic boundary conditions,
for which the Brillouin zone is a finite set of points. 
However, to deal with the continuous
crystal momentum $\bfk$ required to discuss the continuity, 
singularity or analyticity of band energies $\epsilon_n(\bfk)$ 
and Bloch states $\psi_\bfk$, we need to consider infinite crystals.
The symmetry groups of infinite crystals
belong to the category of infinite non-compact groups, 
for which many standard tools of group theory break down.

For example, character theory is no longer available for 
these groups and we use harmonic analysis to build
the group algebra, the regular representation,
the induction of irreducible representations of the crystallographic group
from projective representations of the point groups and the decomposition
of a representation into its irreducible parts.
We deal with magnetic and non-magnetic groups in arbitrary dimensions. 
In the last part of the paper, we discuss Mackey's restriction of 
an induced representation to a subgroup, the tensor product
of induced representations and the symmetric and
antisymmetric squares of induced representations. 
\end{abstract}

\maketitle

\section{Introduction}

Space groups and their representations are essential
tools for the theory of electrons, phonons and magnons in solids, 
as well as for the description of the crystallographic and
magnetic structures of materials. 
Space groups are thoroughly studied and widely used in condensed matter calculations,
many textbooks are devoted to 
them~\cite{Koster,Janssen,Brown-cryst-group,Burns,Schwarzenberger,Opechowski,Birman,Kovalev,Mirman,Hahn,Kim,Dresselhaus,Bradley-10,Melnikov,Szczepanski},
computer packages and web sites (e.g. {\tt{www.cryst.ehu.es}}) describe their properties.
Still, space groups face a fundamental open problem that we want to solve in the present paper. 

The properties of a crystalline sample, which is a finite object,
are most often investigated by representing it either as an infinite  crystal or as
a finite crystal with Born-von K{\`a}rm{\`a}n boundary conditions~\cite{Born-12} linking opposite sides of the crystal
(that we call a \emph{toric crystal} because it is topologically equivalent to a torus).
In solid-state physics, both infinite and toric crystals are used,
in spite of the differences between the two models~\cite{Evarestov-93}.

On the one hand, an infinite crystal has to be used when the wavevector $\bfk$ varies continuously,
for example to study the continuity of an energy band $E_n(\bfk)$ with respect to
$\bfk$ or to know whether the Wannier functions can be
exponentially localized~\cite{BrouderPanati}.
On the other hand, a (finite) toric crystal is used to investigate the symmetry properties
of a crystal, because the symmetry group of a toric crystal is finite and
powerful techniques are available to work with finite groups. However, the Brillouin zone
of a toric crystal is a finite set of points and no continuity or analyticity
questions can be addressed. As a concrete consequence, 
Bloch functions computed at a finite number of points of the Brillouin 
zone pick up random phases from one point to the other
and making these phases ``consistent'' is an additional work
that has not to be done in an infinite crystal. 
Moreover, it is necessary to address the problem of the influence of the size 
of the torus (i.e. the number of unit cells in each direction 
for which the toric crystal is periodic) and to prove
the validity of the thermodynamic limit (i.e. the limit
where the size of the torus tends to infinity) for the observables of interest.

In many papers, even by the most distinguised authors, both points of views are 
intermingled without proper justification. For instance
the landmark paper of solid-state physics by Bouckaert, Smoluchowski and 
Wigner~\cite{Bouckaert-36} uses both results on irreducible representations
of space groups derived for a finite toric crystal~\cite{Seitz-36}
and a continuity argument requiring an infinite crystal.

The present paper aims at providing tools and proofs to deal consistently with 
crystal symmetries and continuous wavevectors, i.e. to work with infinite crystals. 
The main difficulty is that many problems break loose
when we go from finite to infinite groups: an infinite group  can have
infinite-dimensional irreducible representations (as for
the familiar Lorentz group), its collection of irreducible representations can be
so wild that it is not classifiable modulo equivalence~\cite[Thm.~8.F.3]{Bekka-21},  the regular representation 
might not be a (generalized) sum of irreducible representations 
in a \emph{unique} way~\cite[Thm.~1.G.11]{Bekka-21}
and a finite-dimensional representation is
is not necessarily equivalent to a unitary representation~\cite{Segal-50},
to name just a few of them. 

A further difference specific to the case of crystallographic groups 
has to be considered: if $\bfk$ is a vector of the Brillouin zone and
$\bftm$ a lattice vector, we can
define $f(n)=e^{in \bfk\cdot\bftm}$ for any (signed) integer $n$. 
We can see that $f(n)=e^{2\pi i\alpha n}$,
where $\alpha$ is a rational number when the crystallographic group is finite
while $\alpha$ can be an irrational number when the group is infinite. 
In the first case $f(n)$ is periodic, it takes a finite number of values and
each value is reached an infinite number of times. In the second
case $f(n)$ takes an infinite number of different values and each value
is reached only once. This ergodic behavior considerably complicates
the discussion of the characters of irreducible representations, which
become \emph{almost-period} functions in the sense of von Neumann~\cite{Neumann-34}.
As a consequence of these differences between finite and infinite groups, it is not clear 
whether results derived for finite toric crystals are still valid
for infinite crystals.

As we shall see, the main technical difference between finite and infinite groups
is the necessity to use tools from analysis (topology, integrals, distributions, etc.) 
instead of finite sums. This beautiful mixture of 
group theory and functional analysis is called \emph{harmonic analysis}
(see Ref.~\cite{Mackey-80} for a history of the subject).

We not only want to provide a mathematically correct description
of the irreducible representations of infinite crystallographic groups 
in arbitrary dimension, but we also wish to
provide safe tools that solid-state physicists can use to investigate infinite crystals.
We deal with crystallographic groups in arbitrary dimension 
because crystallographic groups of dimensions higher than 3
(i.e. superspace crystallography) are used 
to describe quasicrystals, incommensurately modulated crystals~\cite{Janner-01}
as well as incommensurate magnetic structures~\cite{Rodriguez-24}.
The extension of results on 3-dimensional space groups to arbitrary dimensions is not always
trivial because some papers use the fact that space groups are solvable~\cite{Seitz-36}
and this is no longer true for dimensions $n\ge 4$. 

Crystallographic groups were thoroughly studied by mathematicians because
they  are related to Hilbert's 18th problem and the solution of any Hilbert's problem 
brings eternal fame to its solver. However, irreducible representations of 
crystallographic groups are not well known in the mathematical literature and we have 
recourse to the more general (but highly complex) theory of locally compact groups, 
for which approachable textbooks are now available~\cite{Kaniuth-13,Bekka-21,Gallier-25-I,Gallier-25}.
Still,  the gap between the terminology used in physics and that used in mathematical group theory
(modules, orbifolds, imprimitivity, short exact sequence, Mackey machine, etc.) 
is wide and this paper is partly devoted to the translation of these terms into notions commonly 
used in physics.

Magnetic crystallographic groups in general dimensions received much less attention
and there are very few works on the representation of 
locally compact groups by unitary and antiunitary operators~\cite{Neeb-17}.
Several results are presented at the end of the paper.

In the last section of the paper, we present the generalization
	to crystallographic groups of several useful tools:
	Mackey's restriction of an induced representation to a subgroup, the tensor product
of induced representations and the symmetric and
antisymmetric squares of induced representations.

Irreducible representations are implicitly unitary and often called \emph{irreps}
for short. The Hilbert spaces considered in this paper are implicitly complex and separable.

\section{Crystallographic groups}

A crystal is a pattern of atoms and ions which is regularly repeated in all directions. 
The pattern of atoms can be a single atom, or a molecule that can be as large as a protein. 
In $n$ dimensions, the repetition is ensured by symmetry operations denoted by $\cgop{R}{\bft}$,
consisting of 
a translation by a vector $\bft$ of $\bbR^n$ and a rotation (possibly with an inversion) $R$,
represented~\cite{InfSG1} 
by a real $n\times n$ matrix $M$ such that $M M^T=\id_n$. These operations
act on a point $\bfr$ of $\bbR^n$ by transporting it to $\bfr'=R \bfr + \bft$.
The condition $M M^T=\id_n$ implies that $\cgop{R}{\bft}$ is an isometry:
it conserves the distance between points, which means that it transports
the pattern of atoms without deforming it.

The product of two symmetry operations is defined by
$\cgop{R}{\bft}\cdot \cgop{R'}{\bft'}=\cgop{RR'}{R\bft'+\bft}$, the inverse of 
$\cgop{R}{\bft}$ is $\cgop{R^{-1}}{-R^{-1}\bft}$ and the unit element is 
$\cgop{\un}{0}$.
A crystallographic group $G$ is a group of symmetry operations which is discrete, in the sense
that the distance between copies of the pattern cannot be infinitesimally small
(a physically very desirable condition).

A mathematical analysis of crystallographic groups in arbitrary 
dimension~\cite{Schwarzenberger,Farkas-81,Szczepanski} 
gives the following fundamental results:
\begin{enumerate}
\item  All the elements of a crystallographic groups $G$ can
be written as $\cgop{R}{\bft_R+\bftm}$ where
\begin{eqnarray*}
\bftm &=& \sum_{i=1}^n m_i \bfa_i,
\end{eqnarray*}
with $ \bfa_1,\dots,\bfa_n$ a basis of $\bbR^n$ and all $m_i$ run over $\bbZ$.
\item The vectors $\bft_R$ in $\cgop{R}{\bft_R+\bftm}$ can always
be chosen in the set $C=\{\sum_{i=1}^n x_i \bfa_i\, ;\, 0\le x_i < 1\}$,
which is called the \emph{unit cell} of the lattice. 
\item If we denote by $\Lambda$ the set of all possible $\bftm$,
the group $T=\{\cgop{\un}{\bftm} | \bftm\in\Lambda\}$
is a commutative subgroup of $G$ called the \emph{group of translations} of $G$.
\item The set of $R$ such that $\cgop{R}{\bft_R}$ is in $G$ forms
a finite group, denoted by $G_0$ and called the \emph{point group} of $G$.
\end{enumerate}

\begin{table}
\caption{Number of crystallographic groups for 
dimensions 1 to 6~\cite{Brown-cryst-group,Plesken-00,Cid-01}
(with corrections). (Super)-space groups are classes of crystallographic groups that
are equivalent if and only if they are related by
an isometry $\cgop{R}{\bft}$ such that $\det R=1$. This equivalence
is more physical than  the mathematical one (where $\det R=\pm1$)
because two crystals
which are transformed into one another by a mirror symmetry 
have different physical properties~\cite{Souvignier-03}.
A crystallographic group $G$ is said to be \emph{torsion-free}
(or Bieberbach) if
there is no $g$ in $G$, different from the unit element, such that
the equation $g\bfr=\bfr$ has a solution
(i.e. the only Wyckoff position~\cite{Nespolo-08} is the general one).
The number of torsion-free groups is with respect to 
the crystallographic group classification
(e.g. in dimension 3 there are 10 torsion-free groups among 219
crystallographic groups)}.

\label{table-groupe}
\begin{center}
\bgroup
\def\arraystretch{1.3}
\begin{tabular}{|c|c|c|c|c|c|c|} \hline
Dimension & 1 & 2 & 3 & 4 & 5 & 6\\
\hline 
Crystal systems & 1  & 4 & 7 & 33 & 59 & 251 \\
Bravais lattice types & 1 & 5 & 14 & 64 & 189 & 841\\
Point groups & 2 & 10 & 32 & 227 & 955 & 7103\\
Crystallographic groups  & 2 & 17 & 219 & 4783 & 222018 & 28927915 \\
(Super)-space groups & 2 & 17 & 230 & 4894 & 222097 & 28934967 \\
Torsion-free groups & 1 & 2 & 10 & 74 & 1060 & 38746 \\
\hline
\end{tabular}
\egroup
\vskip 3mm
\end{center}
\end{table}

We can add a few remarks.
The set $\Lambda=\{\sum_{i=1}^n m_i \bfa_i | m_i\in\bbZ\}$ is called
a \emph{Bravais lattice} and the vectors $\bfa_1,\dots,\bfa_n$ are
called \emph{fundamental translations}~\cite[p.~40]{Bradley-10}
and form
a \emph{primitive basis} of $\Lambda$~\cite[p.~296]{Kim}. Bravais
lattices can be classified in various ways~\cite{Pitteri-96,Eick-06}. The most common 
classifications used in crystallography are 
crystal systems and lattice types,
which were enumerated for dimensions $n=1$ to 6 (see Table~\ref{table-groupe}).

The group $T$ of translations of $G$ 
is an invariant (or normal)
subgroup of $G$: for any $g$ in $G$ and any $h$ in $T$ the conjugation $g h g^{-1}$ of
$h$ by $g$ is in $T$. 
The point group $G_0$ of $G$ is a symmetry group of the Bravais lattice: 
for any $R$ in $G_0$ and any Bravais lattice vector $\bftm$, the vector $R \bftm$
is a Bravais lattice vector. The point group $G_0$ of $G$ is generally 
not a subgroup of $G$. If it is, then $G$ is called a \emph{symmorphic} group.
Our analysis will rely on the fact that crystallographic groups and
their subgroups of translation are locally compact (as any discrete group). 
To build the irreducible representations of the infinite group $G$, we start from
the irreducible representations of the infinite commutative group $T$
that we consider now.

\section{Irreducible representations of the group $T$ of translations}
The translation group $T$ of a crystallographic group $G$ has a
very simple structure since, as we shall see, it is isomorphic the commutative
group $\bbZ^n$ made of  $n$ copies of the
group $\bbZ$ of (signed) integers where the group product is
the addition of integers. In spite of this simplicity, the construction of the 
regular representation of $T$, of its group algebra, of its irreducible 
representations and the nature of its characters will require some work
because $T$ is an infinite group.

Another (smaller) difficulty is that, although $T$ is mathematically similar to $\bbZ^n$,
it is physically different from it because the fundamental translations $\bfa_i$ of
the Bravais lattice have definite lengths related to the distances between atoms in the
crystal. As a consequence, the irreducible representations of $T$ will be described
by vectors in the Brillouin zone and the results in the mathematical literature must
be adapted to this situation. We start by defining the Brillouin zone in $n$ dimensions.

\subsection{Irreducible representations of $T$ and the Brillouin zone}
Wigner showed~\cite{Wigner} that quantum physics describes symmetry
by unitary (or anti-unitary) operators $\rho$ acting on some Hilbert space $\calH$.
In the case of finite toric crystals, $\calH$ is a finite-dimensional vector space 
but we now need to deal with possibly infinite-dimensional Hilbert spaces.

Irreducible unitary representations of commutative finite groups are one-dimensional
and the same is true for commutative locally compact infinite groups~\cite[Proposition~3.12]{Gallier-25}
and, being one-dimensional, a representation can be identified
with its character, denoted by $\chi$~\cite[Proposition~3.13]{Gallier-25}.

We now relate the product of elements of $T$ to the addition
of elements of $\bbZ^n$. If $t=\cgop{\un}{\bftm}$ and $t'=\cgop{\un}{\bftm'}$ are elements of $T$,
then $t\cdot t'=\cgop{\un}{\bftm+\bftm'}$,
where $\bftm+\bftm'=\sum_{i=1}^n (m_i+m'_i)\bfa_i$.
We see that the group $T$ with the product $t\cdot t'$ 
is isomorphic to the additive group $\bbZ^n$.
Mathematicians have shown that the 
irreducible representations of $\bbZ^n$ are indexed
by $n$-tuples $(\theta_1,\dots,\theta_n)$ and
take the form~\cite[Corollary~10.11]{Gallier-25-I}
\begin{eqnarray*}
\chi_{(\theta_1,\dots,\theta_n)}\big((m_1,\dots,m_n)\big) &=& e^{\imath(m_1\theta_1+\dots+m_n\theta_n)},
\end{eqnarray*}
where each $\theta_i$ runs over the one-torus $\{\theta_i\in [-\pi,\pi[ \}$ with periodic boundary conditions.

To adapt this result to the physics notation, we rewrite the
representations as $\chi_\bfk(\bftm)=e^{-\imath\bfk\cdot\bftm}$,
where we changed the sign before $\imath$ and where 
$\bfk$  belongs to the (first) Brillouin zone of $T$
that we construct as follows. 

We start by introducing the reciprocal lattice $\Lambda^*$ of $\Lambda$ as the set 
\begin{eqnarray*}
\Lambda^* &=& \{\bfkn = \sum_{i=1}^n n_i \bfb_i | n_i\in \bbZ\},
\end{eqnarray*}
where the reciprocal basis vectors are defined by  $\bfa_i\cdot \bfb_j=2\pi\delta_{ij}$.
By construction, the product $\bfkn\cdot\bftm$ is an integer multiple of $2\pi$
whenever $\bftm$ belongs to $\Lambda$ and $\bfkn$ belongs to $\Lambda^*$.
This can also be interpreted by saying that $\chi_{\bfk+\bfkn}=\chi_\bfk$,
so that two $\bfk$ denote the same irreducible representation if they differ
by an element of $\Lambda^*$. This leads us to identify irreducible representations
with vectors in the Brillouin zone $\BZ$, which is defined as the set of
points closer to the origin of $\Lambda^*$
than to any other point of $\Lambda^*$:
\begin{eqnarray*}
\BZ &=& \{\bfk\in\bbR^n ; ||\bfk|| \le ||\bfk-\bfkn|| \text{ for  all }\bfkn\in\Lambda^*\}.
\end{eqnarray*}
This can also be written
\begin{eqnarray*}
\BZ &=& \{\bfk\in\bbR^n ; -\frac12||\bfkn||^2 \le \bfk\cdot\bfkn \le \frac12||\bfkn||^2 \text{ for  all }\bfkn\in\Lambda^*\}.
\end{eqnarray*}
We observe that $\BZ$ is a convex set and that if $\bfk$ is
in the interior of $\BZ$ (i.e. such that $||\bfk|| < ||\bfk-\bfkn|| $ for all $\bfkn\not=0$),
it is the unique reference to an irreducible representation, whereas
if $\bfk$ lies on the surface of $\BZ$ (i.e. if
$||\bfk|| = ||\bfk-\bfkn|| $ for some $\bfkn\not=0$), then $\bfk$ is on the plane
bissecting the vector $\bfkn$ and it refers to the 
same irreducible representation as $\bfk-\bfkn$
which is also on the surface of $\BZ$, on the plane
bissecting the vector $-\bfkn$. Therefore, both planes
must be identified and we recover the topology of an $n$-dimensional torus
of the mathematical description.

\subsection{The Brillouin zone as a group}
The mathematical literature makes use of the fact that the Brillouin zone is
itself a group~\cite{InfSG2}.
The ``product'' of two elements $\bfk$ and $\bfk'$ of $\BZ$ is 
derived from the product of representations: 
$\chi_\bfk(\bftm)\chi_{\bfk'}(\bftm)=\chi_{\bfk\bullet\bfk'}(\bftm)$ with
$\bfk \bullet \bfk'=\bfk+\bfk'-\bfkn$, where $\bfkn\in\Lambda^*$ is chosen so
that $\bfk+\bfk'-\bfkn$ is in $\BZ$.
This is again a group with an infinite number of elements
but it is now a compact group: the volume $|\BZ|$ of $\BZ$
is finite and  can
be calculated explicitly as $|\BZ|=|\det B|$,
where $B$ is the $n\times n$ matrix whose
columns are $\bfb_1,\dots,\bfb_n$. 

Compact groups such as $\BZ$ have almost
all the nice properties of finite groups~\cite{Serre-group}.
The main difference is that 
\begin{eqnarray*}
\frac{1}{|G|}\sum_{g\in G}  f(g) &\text{is\,\,replaced\,\,by}&
\frac{1}{|\BZ|} \int_\BZ f(\bfk) d\bfk. 
\end{eqnarray*}

The irreducible representations of 
the compact commutative group $\BZ$
are indexed by $\bftm\in\Lambda$ and
their values are
$\chi_{\bftm}(\bfk)=e^{-\imath\bfk\cdot\bftm}$.
This relation between the groups $T$ and $\BZ$
leads to a generalization of the Fourier transform, as
we shall see below.

In the same way as the Bravais lattice $\Lambda$ can be 
considered as a group whose dual is the group $\BZ$,
the reciprocal lattice $\Lambda^*$ can be 
seen as a group and its (compact) dual group is the unit cell $C$
with periodic boundary conditions. If $A$ denotes the $n\times n$ matrix whose
columns are $\bfa_1,\dots,\bfa_n$, then
the volume $|C|$ of the unit cell of $\Lambda$ is $|C|=|\det A|$
and the relation $A^TB=(2\pi)\id$ between $A$ and $B$ implies
$\det A\det B=(2\pi)^n$ and
$|\BZ| = (2\pi)^n/|C|$.

\subsection{Group algebra}

When $G$ is a finite group, it is very useful to define the \emph{group algebra}
$A=\{\sum_g a_g g\}$, where all $a_g \in \bbC$.
This algebra enables us to define the sum of two elements
$a=\sum_g a_g g$ and $b=\sum_g b_g g$ of $A$ by $a+b=\sum_g (a_g +b_g) $
and the product of an element of $a$ by a complex  number $\lambda$
as $\lambda a=\sum_g (\lambda a_g) g$. The group product
gives us an algebra product:
\begin{eqnarray*}
a \cdot b &=& \big(\sum_{g_1} a_{g_1} g_1\big)\cdot \big(\sum_{g_2} b_{g_2} g_2\big)
= \sum_{g_1 g_2 } a_{g_1} b_{g_2} (g_1\cdot  g_2)
\\&=&
\sum_g \big(\sum_{h} a_{gh^{-1}} b_{h} \big) g.
\end{eqnarray*}
The group algebra enables us to define the regular representation,
projection operators, irreducible idempotents, etc.~\cite{LudwigFalter}.
Moreover, a representation $(V,\rho)$ of $G$ becomes a 
representation~\cite{InfSG3}
of $A$ defined by $\rho(a)=\sum_g a_g \rho(g)$.

From the mathematical point of view, it is more
convenient to replace the algebra $A$ whose elements are $a=\sum_g a_g g$
by the algebra of functions $f:G\to \bbC$ defined by
$f(g)=a_g$. The sum of elements of the algebra $A$ and its product by a complex number $\lambda$
become the sum of the corresponding functions and its product by $\lambda$.
The product $a \cdot b$ in the algebra $A$
becomes the convolution product of functions
\begin{eqnarray}
(f \star f')(g) &=& \sum_{h\in G} f(gh^{-1}) f'(h).
\label{convprod}
\end{eqnarray}
Finally, the representation $(V,\rho)$ of $G$ induces
the representation $\rho(f)=\sum_g f(g) \rho(g)$.

For an infinite group such as $T$ or $G$, the infinite sum in the convolution
product Eq.~\eqref{convprod} is not always well defined.
To deal with this problem we define, for any  function $f:G\to\bbC$, its
(possibly infinite) $p$-norm as 
\begin{eqnarray}
||f||_p &=& \Big(\sum_{g\in G} |f(g)|^p\Big)^{1/p}, \label{deffp}
\end{eqnarray}
that we extend to $||f||_\infty=\sup_{g\in G} |f(g)|$.
The \emph{sequence space} $\ell^p(G)$ is the space of functions 
such that $||f||_p$ is finite. It is a Banach space
(i.e. the vector space is complete for the topology
induced by the metric $||\cdot||_p$).
In the case of crystallographic groups $G$, we have
$\ell^p(G)\subset \ell^q(G)$ whenever $1\le p\le q\le \infty$~\cite{Miamee-91}.

Two sequence spaces will be particularly useful: $\ell^2(G)$
because it is a Hilbert space and $\ell^1(G)$ because it can
be considered as the group algebra of $G$.

The space $\ell^2(G)$ is a Hilbert space~\cite[Thm.~5.45]{Gallier-25-I} 
when equipped with the scalar product
\begin{eqnarray*}
\langle f,f' \rangle  &=& \sum_{g\in G} \overline{f(g)}  f'(g).
\end{eqnarray*}

We consider now the Brillouin zone group $\BZ$
and functions from $\BZ$ to $\bbC$. 
Since the Brillouin zone is seen as an $n$-dimensional torus,
these functions are periodic:
$f(\bfk+\bfkn)=f(\bfk)$ for every $\bfk$ of $\BZ$ and 
every $\bfkn$ of the reciprocal lattice. 
Bloch states $\psi_{n,\bfk}(\bfr)$ are examples of these functions
(for each $\bfr$). The $p$-seminorm of a measurable function $f$ on $\Omega$
is
\begin{eqnarray*}
||f||_p &=& \Big(\frac{1}{|\BZ|} \int_\BZ d\bfk |f(\bfk)|^p\Big)^{1/p},
\end{eqnarray*}
and the function space $L^p(\BZ)$ is the set of functions
for which $||f||_p$ is finite
(more precisely the quotient of this set by the set of
functions that are zero almost everywhere). 
Note that we have now
$L^q(G)\subset L^p(G)$ whenever $p\le q\le \infty$
(the inclusion is reversed with respect to $\ell^p(G)$)~\cite{Miamee-91}.
Again, $L^2(G)$ is a Hilbert space for the scalar product
\begin{eqnarray*}
\langle f,f' \rangle  &=& \frac{1}{|\BZ|} \int_\BZ d\bfk  \overline{f(\bfk)}  f'(\bfk).
\end{eqnarray*}

\subsection{Fourier transformation of $T$ and $\BZ$}
The Fourier transformation can be generalized to
commutative locally compact groups~\cite[Chapter~6]{Gallier-25-I}.
In our case, it creates a link between the translation group $T$ and
the Brillouin zone group $\BZ$. 
If $f : T \to \bbC$ is absolutely summable,
its Fourier transformation is 
\begin{eqnarray}
\tildef(\bfk)  &=& \calF(f)(\bfk)=\sum_{\bftm\in\Lambda} e^{-\imath\bfk\cdot\bftm} f(\bftm),
\label{defFourierT}
\end{eqnarray}
where we write $f(\bftm)$ for $f\big(\cgop{\un}{\bftm}\big)$. 
When the sum is well defined, $\tildef(\bfk+\bfkn)=\tildef(\bfk) $
and  $\tildef(\bfk)$ is indeed a function from $\BZ$ to $\bbC$.   

The inverse Fourier transformation is
\begin{eqnarray*}
\calF^{-1}(\tildef)(\bftn) &=&
\frac{1}{|\BZ|} \int_\BZ d\bfk e^{\imath\bfk\cdot\bftn}\tildef(\bfk),
\end{eqnarray*}
so that $(\calF^{-1}\circ\calF)(f)=f$. 
In particular, if we consider the function $f(\bftm)=\delta_{\bftm,0}$ we obtain
a useful identity to investigate the band-structure or phonons
of infinite crystals:
\begin{eqnarray}
\frac{1}{|\BZ|} \int_{\BZ} e^{\imath\bfk\cdot\bftn} d\bfk&=&
\delta_{\bfn,0}.
\end{eqnarray}
By working with the reciprocal lattice group $\Lambda^*$ and the unit cell group $C$
we similarly obtain:
\begin{eqnarray}
\frac{1}{|C|}\int_C d\bfr e^{\imath\bfk_\bfm\cdot\bfr} &=& \delta_{\bfm,0}.
\end{eqnarray}

Various function spaces are useful to work with Fourier transformation~\cite[Chapter~6]{Gallier-25-I}.
We only mention $\ell^2(T)$ because of the importance of Parseval's theorem: $f$ belongs to $\ell^2(T)$ if 
and only if $\calF(f)$ belongs to $L^2(\BZ)$ and they are related by $||f||_2=||\calF(f)||_2$. 

A last application of the Fourier transformation 
is the relation between two ways to construct a 
periodic function from a general rapidly decreasing smooth function
$f:\bbR^n\to\bbC$, either by Fourier transform or
by translation along all vectors of $\Lambda$~\cite[Section~6.9]{Gallier-25-I}:
\begin{eqnarray*}
\sum_{\bfkn\in\Lambda^*} e^{\imath\bfkn\cdot\bfr} \int_{\bbR^n} d\bfr e^{-\imath\bfkn\cdot\bfr} f(\bfr)
&=& |C| \sum_{\bftm\in\Lambda} f(\bfr+\bftm)
\end{eqnarray*}
This is the famous Poisson formula, where the factor $|C|$ comes from the fact
that $d\bfr$  is a volume element in the Fourier transform.
This formula is initially derived for rapidly decreasing smooth functions $f$
and then extended by duality to tempered distributions. When $f$ is the Dirac delta distribution, 
we obtain a very useful formula of solid-state physics calculations~\cite{BrouderRossano}:
\begin{eqnarray}
\sum_{\bfkn\in\Lambda^*} e^{\imath\bfkn\cdot\bfr}
&=& |C| \sum_{\bftm\in\Lambda} \delta(\bfr+\bftm),
\label{PoissonLambdastar}
\end{eqnarray}
and its companion
\begin{eqnarray}
\sum_{\bftm\in\Lambda} e^{\imath\bfk\cdot\bftm}
&=& |\BZ| \sum_{\bfkn\in\Lambda^*} \delta(\bfk+\bfkn).
\label{PoissonLambda}
\end{eqnarray}

\section{Representations of crystallographic groups}
In this section we describe the general properties of 
the unitary representations of crystallographic groups
and we start with the description of some properties of
crystallographic groups.

\subsection{Properties of crystallographic groups}
We first consider the conjugacy classes in crystallographic groups.
If $g=\cgop{R}{\bft}$ and $g'=\cgop{R'}{\bft'}$
are two operations of a crystallographic group $G$, the conjugation of
$g'$ by $g$ is
\begin{eqnarray*}
g\cdot g'\cdot g^{-1} &=&\cgop{R}{\bft}\cdot \cgop{R'}{\bft'} \cdot \cgop{R^{-1}}{-R^{-1}\bft}
\\&=&
\cgop{RR'R^{-1}}{-RR'R^{-1}\bft+R\bft'+\bft}.
\end{eqnarray*}
We first check that if $g'=\cgop{\id}{\bftm}$ is in $T$, then
$g\cdot g'\cdot g^{-1}=\cgop{\id}{R\bftm}$ is in $T$ and $T$ is an invariant subgroup of $G$.
As a consequence, there is a quotient group $G/T$. 
To describe it, we determine when $g=\cgop{R}{\bft_R+\bftm}$ and $g'=\cgop{R'}{\bft_{R'} +\bftm'}$ belong to the same coset in $G/T$.
We have $g'g^{-1}= \{R'R^{-1}, -R'R^{-1}(\bft_{R} +\bftm)+ \bft_{R'}+\bftm'$; since $\bftm$
and $-\bftm'$ are in $\Lambda,$, it follows that $g'g^{-1}$ is in $T$ if and only if $R'=R$.
In other words, two elements of $G$ are in the same coset
iff they have the same $R$ and their translation parts differ by 
an element of $\Lambda$.
Therefore, the number of cosets is equal to the number of elements of $G_0$,
which is finite. In other words, the index $[G{:}T]$ of $T$ in $G$ is finite.


\subsection{The group algebra}

For any crystallographic group $G$, the 
space $\ell^1(G)$ equipped with the convolution product
is an algebra~\cite{InfSG4}, called the \emph{group algebra of $G$}.
This algebra has a unit: $\delta_{g,\unit}$, where $\unit$ is the unit of $G$. 

It will be useful to define the space $\ell^\infty(G)$ of 
\emph{bounded functions} from $G$ to $\bbC$, which is the set
of functions $\chi: G\to\bbC$ for which there is a number
$M$ such that $|\chi(g)| < M$ for all $g\in G$.
The importance of this set comes from the fact that the
matrix coefficients of unitary representations of $G$ belong to $\ell^\infty(G)$.
Boundedness implies that there is a well-defined bilinear coupling between $f$ in $\ell^1(G)$ and 
$\chi$ in $\ell^\infty(G)$ by~\cite[Def.~5.21]{Gallier-25-I}:
\begin{eqnarray*}
(f,\chi) &=& \sum_{g\in G} f(g)\chi(g),
\end{eqnarray*}
which is non-degenerate: if $(f,\chi)=0$ for all $\chi$
then $f=0$ and if $(f,\chi)=0$ for all $f$
then $\chi=0$~\cite[Thm.~5.51]{Gallier-25-I}.
This coupling also defines a norm-preserving isomorphism 
between $\ell^\infty(G)$ and the dual of $\ell^1(G)$
as well as a norm-preserving injective map
between $\ell^1(G)$ and the dual of 
$\ell^\infty(G)$~\cite[Thm.~5.51]{Gallier-25-I}.

The group algebra of the Brillouin zone group $\Omega$ is more complicated.
A natural candidate for the group algebra seems to be  $L^1(\BZ)$ 
with the convolution product
\begin{eqnarray*}
(f \star f')(\bfk) &=&
\frac{1}{|\BZ|} \int_\BZ d\bfq f(\bfk-\bfq) f'(\bfq).
\end{eqnarray*}
However, the unit of this algebra would be 
$|\Omega|\delta(\bfk)$,
where $\delta(\bfk)$ is the Dirac delta distribution, 
which is not in $L^1(\BZ)$.

Therefore, we extend the group algebra to
the space of measures $M^1(\BZ)$ (see~\cite[Def.~7.22]{Gallier-25-I} for a
detailed definition). This extension enables us to include 
the Dirac delta distribution and the unit of the algebra
$M^1(\BZ)$.
A maybe more familiar way to define $M^1(\BZ)$ is to start
from $\calC(\BZ)$, the space of functions from 
$\BZ$ to $\bbC$, continuous for the topology of $\BZ$ defined
from the standard metric $||\cdot||$ by 
$d(\bfk,\bfk')=\min_{\bfkm\in\Lambda^*}||\bfk-\bfk'+\bfkm||$.
The space $M^1(\BZ)$ can be seen as the dual $\calC(\BZ)'$
of $\calC(\BZ)$. This dual is the space of continuous linear forms 
$\lambda: \calC(\BZ) \to \bbC$. 
It is a Banach space for the norm
$||\lambda||=\sup\{|\lambda(f), f\in\calC(\BZ), ||f||\le 1 \}$
(this is a consequence of Ref.~\cite[Prop.~7.5]{Gallier-25-I}
and $\calC_0(\BZ)=\calC(\BZ)$ when
$\BZ$ is compact~\cite[p.~41]{Gallier-25-I}).

\subsection{The regular representation}
\label{regularepsect}
For a finite group $G$, the group algebra
is made of elements $a=\sum_g a_g g$ with $a_g\in\bbC$.
The \emph{left regular representation} of $h\in G$ is defined on this algebra by 
\begin{eqnarray*}
\lambda_G(h) \big( \sum_g a_g g \big)&=& \sum_g a_g (h\cdot g)
=\sum_g a_{h^{-1}\cdot g}\, g.
\end{eqnarray*}
If we consider now the group algebra seen as
the algebra of functions $f:G\to\bbC$,
the (left) regular representation of the group algebra
transforms the function $f$
into the function $\lambda_G(h)f$ defined by
$\big(\lambda_G(h)f)(g)=f(h^{-1}\cdot g)$,
where $g$ and $h$ are in $G$.

In the case of a crystallographic group $G$, we can no longer use 
arbitrary functions because $G$ is infinite and the correct space for the regular representation
turns out to be $\ell^1(G)$ acting on the Hilbert space $\ell^2(G)$. For any 
$f\in \ell^2(G)$ we can write again~\cite[Section~3.3]{Gallier-25}
$\big(\lambda_G(h)f)(g)=f(h^{-1}\cdot g)$.
So we can extend the left regular representation
from $G$ to the group algebra $\ell^1(G)$ and a function $f$ of $\ell^1(G)$ 
is represented on the Hilbert space $\ell^2(G)$
by the convolution product~\cite[Def.~3.15]{Gallier-25}
\begin{eqnarray*}
\lambda_G(f) v &=& f\star v,
\end{eqnarray*}
for every $v\in \ell^2(G)$.

\subsection{Irreducible representations}
\label{irrepsect}
We must be more precise in the definition of irreducible
representations of infinite groups.
A unitary representation $(\calH,\rho)$ of a locally compact group $G$
is said to be \emph{reducible} if there is a non-zero
closed proper subspace $V$ of $\calH$
such that $\rho(g)V\subset V$ for every $g\in G$
(if $V$ is not closed, it can be replaced by its
closure~\cite[Prop.~3.5]{Gallier-25}).
A representation is \emph{irreducible} if it is not reducible. 
Note that Schur's lemma is still valid for unitary representations of 
locally compact groups~\cite[Prop.1.A.11]{Bekka-21}.

Two irreducible representations $(\calH_1,\rho_1)$ and
$(\calH_2,\rho_2)$ are considered equivalent if
there exists a unitary operator $U$ from $\calH_1$
to $\calH_2$
such that $\rho_1(g)=U^\dagger \rho_2(g) U$ for all $g\in G$.
We denote by $\hatG$ the set of equivalence classes of
irreducible representations of $G$.

Irreducible representations of locally compact groups can be of 
type I, II or III~\cite{Bekka-21}. The precise definition of
these three types would bring us too far and suffice it to say that
representations of type I are relatively nice while those of type II and III can be quite wild.
Elmar Thoma proved a theorem~\cite{Thoma-68} 
that turns out to be crucial for our work:
a discrete group $G$ has a commutative 
invariant subgroup $T$ with a finite index if and only if
all unitary representations of $G$ are of type I;
in this case, all irreducible unitary representations of $G$
are finite dimensional and their dimensions
cannot be larger than the index. 

Crystallographic groups $G$ correspond to the 
hypothesis of Thoma's theorem because they
are discrete groups with a commutative
invariant subgroup (the group $T$ of translations)
and the index of $T$ is the number of elements
$|G_0|$ of the point group $G_0$ of $G$. 
Since this number is finite, all representations
of $G$ are of type I and the irreducible
representations of $G$ have dimension 
smaller or equal to $|G_0|$. 
This result will enable us to link the representations
of infinite crystals to those of finite toric crystals.

The second central result we need is that,
when all representations of a group are of type I,
there is a canonical decomposition into irreducible
representations~\cite[Thm.~6.D.7]{Bekka-21}.
We shall see that any irreducible 
representation $\rho$ of a crystallographic group $G$ is entirely
determined by a vector $\bfk\in \IBZ$ and
a projective representation $\tau$ of $G_0(\bfk)$
and we denote it $\rho_{\bfk\tau}$ instead of $\rho$.

With this notation, for every representation $(\calH,\sigma)$
of $G$ in a (separable) Hilbert space $\calH$,
there exist a \emph{unique} set of (possibly infinite) integers $\{m_{\bfk,\tau}\}$
and a positive measure $\mu$ on $\IBZ$
(i. e. the kernel of a positive distribution)
such that:
\begin{eqnarray}
\sigma(g) &=& \int^\oplus_{\IBZ} \mu(\bfk) d\bfk
\bigoplus_{\tau} m_{\bfk,\tau} \rho_{\bfk,\tau}(g)  ,
\label{sumirrep}
\end{eqnarray}
where $\IBZ$ is the irreducible Brillouin zone $\BZ/G_0$,
$\tau$ runs over the small representations of $G(\bfk)$
(or equivalently the irreducible projective representations of $G_0(\bfk)$),
$\rho_{\bfk,\tau}$ is the representation of $G$ induced from 
the representation $\tau$ and 
$m_{\bfk,\tau}$ is the number of times the representation $\rho_{\bfk,\tau}$
appears in $\sigma$. 
The notation $\bigoplus_{\tau}$ and $\int^\oplus$ stands for
direct sum and direct integral. 

We can separate the translation part and the rotation part in the usual way:
\begin{eqnarray*}
\sigma\big(\cgop{R}{\bft_R+\bftm}\big) &=& \int^\oplus_{\IBZ} \mu(\bfk) d\bfk e^{-\imath\bfk\cdot\bftm }
\bigoplus_{\tau} m_{\bfk,\tau} \rho_{\bfk,\tau}\big(\cgop{R}{\bft_R}\big).
\end{eqnarray*}

This decomposition is unique. 
Note that when a representation is not of type
I, several decompositions exist with inequivalent 
irreducible representations~\cite[Thm.~1.G.11]{Bekka-21},
a situation that would be very awkward for quantum
physics applications.

\subsection{Decomposition of the regular represention
and Plancherel theorem}
As an example of the decomposition an infinite dimensional
representation into irreducible representations, we discuss
the decomposition of the regular representation introduced 
in section~\ref{regularepsect}.

For any $f\in\ell^1(G)$ and any irreducible representation
$\sigma_{\bfk\tau}$ we define the matrix
\begin{eqnarray*}
\sigma_{\bfk\tau}(f) &=& \sum_{g\in G} f(g) \sigma_{\bfk\tau}(g),
\end{eqnarray*}
and its Hilbert-Schmidt norm $||\sigma_{\bfk\tau}(f)||$ by
$||\sigma_{\bfk\tau}(f)||^2 = 
\tr \big(\overline{\sigma_{\bfk\tau}(f)}\sigma_{\bfk\tau}(f)\big)$.

Then, \cite[Thm.~4.4]{Kleppner-72} gives us the Plancherel formula
\begin{eqnarray*}
\sum_{g\in G} |f(g)|^2 &=& 
\frac{1}{|\IBZ|}
\int_{\IBZ} d\bfk \sum_\tau \dim\tau\,||\sigma_{\bfk\tau}(f)||^2,
\end{eqnarray*}
where $\tau$ runs over the projective representations of $G_0(\bfk)$,
$|\IBZ|=|\BZ|/|G_0|$ is the volume of the irreducible
Brillouin zone
and $d\bfk$ is the Lebesgue measure.

Correspondingly, the regular representation $\lambda_G$ of $G$ has
the following decomposition
\begin{eqnarray*}
\lambda_G(g) &=&  \frac{1}{|\IBZ|}
\int_{\IBZ} d\bfk \bigoplus_\tau \big(\sigma_{\bfk\tau}(g)\otimes \id_{\dim\tau}\big).
\end{eqnarray*}

\section{Constructing irreps of crystallographic groups}
In this section we shall build the irreducible representations of 
an arbitrary crystallographic group $G$ in dimension $n$.

The textbook construction of the irreducible representations of finite space groups
follows the following steps~\cite{Kim,Bradley-10} 
\begin{enumerate}[(i)]
\item $*$ Construct all irreps $\chi_\bfk$ of $T$
\item Group the irreps of $T$ into different orbits of $\bfk$ 
and choose one member of each orbit.
\item Determine the respective little group $G(\bfk)$ for each selected $\bfk$ of $T$
\item $*$ Find the small representations $\rho$ of the little group $G(\bfk)$ by
using the projective representations of $G(\bfk)/T$.
\item $*$ Check that the irreps of $G$ induced from the small irreps $\rho$ of $G(\bfk)$
are irreducible irreps of $G$
\item $*$ Check that the set of irreps of $G$ thus constructed is complete
\end{enumerate}
The star indicates steps where the proof must be modified to deal
with infinite groups. 
We already carried out step (i) when we discussed the 
group $T$ and the Brillouin zone. For an analog of step (iv) we must first
prove that all irreducible are finite dimensional. This was done in section~\ref{irrepsect}. 

To prove step (v) the finite group case uses
the fact that a representation $(\calH,\rho)$ is 
irreducible iff~\cite{CornwellI}
\begin{eqnarray}
\sum_{g\in G} |\chi(g)|^2 &=& |G|,
\label{sumgchideux}
\end{eqnarray}
where $\chi(g)=\tr\big(\rho(g))$.

The final step (vi) is obtained by using the fact that a
collection of inequivalent irreducible representations $\{(\calH_i,\rho_i)\}_{i\in I}$
is complete iff
\begin{eqnarray}
\sum_{i\in I} \big(\dim\calH_i\big)^2 &=& |G|.
\label{sumidimdeux}
\end{eqnarray}
In crystallographic groups $|G|=\infty$ and
criteria~\eqref{sumgchideux} and \eqref{sumidimdeux} are obviously no 
longer valid. They must be replaced by deep results from noncommutative
harmonic analysis that we present now.

The general construction of irreducible representations of
locally compact group (the so-called \emph{Mackey machine})
is extremely involved~\cite{Folland-95} and detailed examples
are often restricted either to the case of semi-direct products of groups
(e.g. symmorphic crystallographic groups) or to torsion-free crystallographic
groups~\cite{Ramras-14} (i.e. groups where no 
symmetry operations has a fixed point). 
It is only recently that Jean Gallier and Jocelyn Quaintance 
wrote a book~\cite{Gallier-25}, readable by a (hard-working) non-expert, 
where a construction sufficiently general to include crystallographic groups is
described in detail. We now follow their presentation~\cite[Chapter~7]{Gallier-25} 
and adapt it to our problem. Note that some of the tools 
can have an independent interest for solid-state physics,
for example the projection-valued measures.

\subsection{Projection-valued measures}
We first recall that a measure is a way to quantify the size of sets.
Intuitively, the measure
of a group of disjoint objects is the sum of the measures of each object
and the measure of nothing is zero.
In mathematical parlance, if $X$ is a set and $\Sigma$ a collection of 
subsets~\cite{InfSG5} 
of $X$, then $\mu: \Sigma\to [0,+\infty]$ is a measure if:
i) $\mu(\emptyset)=0$ and ii) if $(E_i)_{i\in\bbN}$ is a (countable) collection
of disjoint sets in $\Sigma$ (i.e. $E_i\cap E_j=\emptyset$ if $i\not=j$), then
\begin{eqnarray*}
\mu\Big(\bigcup_{i\in I}E_i\Big) &=& \sum_{i\in I} \mu(E_i).
\end{eqnarray*}
Besides, an orthogonal projection on a Hilbert space $\calH$
is a linear map $P:\calH\to\calH$ such that
$P^2=P$ and $P^\dagger=P$.

From these definitions the following concept of 
a \emph{projection-valued measure} was created~\cite[Def.~2.22]{Gallier-25}.
\begin{dfn}
Let $\calH$ denote a Hibert space, $X$ a set and
$\Sigma$ a collection of subsets of $X$.
A \emph{projection-valued measure} $P$ is a map from $\Sigma$
to the set of orthogonal projections on $\calH$ satisfying
\begin{enumerate}
\item $P(\emptyset)=0$ and $P(X)=\id$
\item  For all countable collection $\{E_i\}_{i\in I}$ of pairwise 
disjoint sets in $\Sigma$ and for every $v\in\calH$
\begin{eqnarray*}
P\Big(\bigcup_{i\in I}E_i\Big)v &=& \sum_{i\in I} P(E_i)v.
\end{eqnarray*}
\item For any $E_1$ and $E_2$  in $\Sigma$
\begin{eqnarray*}
P(E_1\cap E_2) &=& P(E_1)P(E_2).
\end{eqnarray*}
\end{enumerate}
\end{dfn}

Familiar examples of projection-valued measures are the spectral
projections onto the eigenspaces corresponding to a set of eigenvalues.
In quantum physics, the projection valued-measured corresponding to
the Hamiltonian $H$ and the set of 
energies $E=[e_{\min},e_{\max}]$ is usually written
\begin{eqnarray*}
P(E) &=& \sum_{e_{\min} \le e_n \le e_{\max}} |\phi_n\rangle\langle\phi_n|,
\end{eqnarray*}
where $|\phi_n\rangle$ is the eigenstate of $H$ with energy $e_n$.
It is often convenient to write this projection-valued measure as
\begin{eqnarray*}
P(E) &=& \frac{1}{2\pi i} \int_\gamma \frac{dz}{z-H},
\end{eqnarray*}
(where $\gamma$ is a contour around $E$ separating
$E$ from the rest of the spectrum of $H$.

Projection-valued measures were also used to investigate the localizability of
quantum mechanical systems~\cite{Wightman-62,Grigore-93}
or for the construction of maximally localized Wannier functions~\cite{BrouderPanati}
through the use of composite bands~\cite{Blount,Kohn73}.
Projection-valued measures describe pure quantum states
and their generalization to mixed states 
(Positive Operator-Valued Measures)  have become
a central tool for the investigation of
quantum systems~\cite{Holevo,Mauro}.

This lengthy introduction allows us to write the following 
important formula, which is valid for any unitary representation 
of $G$~\cite[Prop.~7.1]{Gallier-25}:
Let $G$ be a crystallographic group and $\BZ$ its Brillouin zone.
For any unitary representation $(\calH,\sigma)$ of $G$, there
is a unique projection-valued measure $P$ such that,
in the standard mathematical notation
\begin{eqnarray}
\sigma\big((\id,\bftn)\big) &=& \frac{1}{|\BZ|}\int_\BZ e^{-\imath\bfk\cdot\bftn} dP(\bfk).
\label{dPk}
\end{eqnarray}
Since this notation is not very familiar in physics, we translate it into
the notion of projection-valued distribution, which is the usual
way to describe quantum fields~\cite{Wightman}. 
We use distributions because we need representations
$\sigma$ corresponding to a single $\bfk_0$
and we write $P(\bfk)d\bfk$ for $dP(\bfk)$,
where, for any $\bfv$ in $\calH$, 
$\langle\bfv|P(\bfk)\bfv\rangle$ is the kernel of a distribution,
and, for any smooth function $f$ on $\Omega$, 
$P(f)=\int f(\bfk)P(\bfk)d\bfk$ is a bounded operator on $\calH$.
With this notation we rewrite Eq.~\eqref{dPk} as
\begin{eqnarray}
\sigma\big((\id,\bftn)\big) &=& \frac{1}{|\BZ|}\int_\BZ e^{-\imath\bfk\cdot\bftn} P(\bfk)  d\bfk.
\end{eqnarray}

Moreover, these projections are covariant:
\begin{eqnarray}
\sigma\big((R,\bftn)\big) P(\bfk) \sigma\big((R,\bftn)\big)^{-1} &=& P(R^{-1}\bfk).
\end{eqnarray}
Finally, if $\sigma$ is irreducible, then for each $\bfk$ in $\BZ$
and each set orbit $E_\bfk=\{R\bfk; R\in G_0\}$ 
either $P(E_\bfk)=\id$ or  $P(E_\bfk)=0$.

Projections in $L^1(G)$ are discussed in~\cite[Section~7.2]{Kaniuth-13}.

\subsection{Action of $G$ on $\BZ$}
To be able to use the mathematical results that we need,
we must show that the (left) action of the crystallographic group
$G$ on the Brillouin zone (or more precisely on the
characters of $T$) is \emph{nice enough}. 
We first describe this action and in the next section we
define what we mean by ``nice enough''.

The left action of an element $g$ of $G$ on 
a character $\chi$ of $T$ is the character 
$g\triangleright \chi$ of $T$ defined by
$(g\triangleright \chi)(t)=\chi(g^{-1}t g)$.
We need $g^{-1}t g$ and not $gtg^{-1}$ to ensure that
$g'\triangleright(g\triangleright \chi)=(g'g)\triangleright\chi$. 
For a crystallographic group operation, we have
\begin{eqnarray*}
\big(\cgop{R}{\bft}\triangleright \chi_{\bfk}\big)\big(\cgop{\un}{\bftm}\big) &=&
\chi_{\bfk}\Big(\cgop{R^{-1}}{-R^{-1}\bft}\cdot \cgop{\un}{\bftm} \cdot \cgop{R}{\bft}\Big)
\\&=&
\chi_{\bfk}\big(\cgop{\un}{R^{-1}\bftm} \big).
\end{eqnarray*}
But 
\begin{eqnarray*}
\chi_{\bfk}\big(\cgop{\un}{R^{-1}\bftm} \big) &=& 
e^{-\imath \bfk\cdot (R^{-1}\bftm)} = e^{-\imath (R\bfk)\cdot\bftm}
=\chi_{R\bfk} \big(\cgop{\un}{\bftm} \big).
\end{eqnarray*}
In other words $\cgop{R}{\bft}\triangleright \chi_{\bfk}=\chi_{R\bfk}$.
In solid-state physics, we rather consider the related action of $G$ on $\BZ$
defined by $\cgop{R}{\bft}\triangleright \bfk=R\bfk$.
This leads us to the following familiar concepts of the theory of space groups: 
\begin{enumerate}
\item the \emph{little group~\cite{InfSG6} of $\bfk$}, that we denote by $G(\bfk)=\{\cgop{R}{\bft}\in G; R\bfk =\bfk+\bfkn
\text{ for some } \bfkn\in\Lambda^*\}$
\item the \emph{star of $\bfk$}, also called the \emph{orbit of $\bfk$}, which is the set
$\calO_\bfk=\{ R\bfk; \text{ for all } \cgop{R}{\bft}\in G\}$, also denoted by $\star \bfk$.
\end{enumerate}
In the mathematical literature, the little group and the orbit are built by acting
on irreducible representations of $T$ instead of on $\BZ$
and the condition is $R\bfk =\bfk$ instead of $R\bfk =\bfk+\bfkn$
but the two are equivalent when $\BZ$ is assumed to be periodic. 

The general theorem asserts that, if the action of $G$ on $\BZ$ is nice enough,
every irreducible representation of $G$ arises as an induced representation
of some irreducible representation of $G(\bfk)$ for some $\bfk\in\BZ$. 
The technical translation of  ``nice enough'' is that the action of
$G$ on $\BZ$ must be \emph{regular}. 
A frustrating aspect of the application of mathematical theorems to physics is
that the hypotheses of the theorems must be checked. 
This is the purpose of the next section.

\subsection{Regularity of the action}
The action of a crystallographic group $G$ on the Brillouin zone $\BZ$
is regular if the following
two conditions hold~\cite[Def.~7.6]{Gallier-25}:
\begin{enumerate}
\item The space of orbits of the action of $G$ on $\BZ$ is countably separated: 
there is a countable family $(E_j)$ of 
$G$-invariant Borel subsets of $\BZ$ such that each orbit $\calO$
is the intersection of all $E_j$ that contain $\calO$:
\begin{eqnarray}
\calO &=& \bigcap\{E_j; \calO\subset E_j\}
\label{calO}
\end{eqnarray}
\item For every $\bfk\in\BZ$, the map $G/G(\bfk)\to\calO_\bfk$ given
by $g \cdot G(\bfk)\mapsto g\triangleright\bfk$ is a homeomorphism.
\end{enumerate}
The purpose of this section is to show that the action of
a crystallographic group $G$ on its Brillouin zone $\BZ$ 
is regular. But we first give a familiar example of an
ergodic (i.e. non-regular) action.

\subsubsection{Regularity versus ergodicity}
The action of an infinite group can
have an unexpected behavior, called ergodicity.
A  simple example of ergodicity is given by the action
of the additive group of signed integers $\bbZ$ (seen as a one-dimensional crystal) on the unit circle 
$C$ by 
\begin{eqnarray*}
m\triangleright \left(\begin{array}{c}\cos\theta\\\sin\theta\end{array}\right) &=& 
\left(\begin{array}{c}\cos(\theta+2\pi m\alpha)\\\sin(\theta+2\pi m\alpha)\end{array}\right),
\end{eqnarray*}
where $m\in\bbZ$.
If $\alpha$ is a rational number $\alpha=p/q$, then the orbit of a point $M$ of $C$ under the action
of $\bbZ$ is a (finite) set of $q$ points (assuming $p$ and $q$ to be relatively prime)
and the action of $\bbZ$ on $C$ is regular.
If $\alpha$ is irrational, then the orbit of any point $M$ of $C$ under the action of $\bbZ$
is an infinite set of points which is dense in $C$: for any point $M'$ of $C$
and any $\epsilon>0$, there is a $m\in\bbZ$ such that $||m\triangleright M-M'|| <\epsilon$. 
Kirillov~\cite[p.~118]{Kirillov} gives other simple examples of groups with ergodic action.

Such an ergodic behavior can be very desirable, for instance
in statistical physics~\cite{Mackey-74}, but this 
not what we need in crystallography because we would not
be able to define the irreducible Brillouin zone, as we shall see.
Therefore, we prove now that the action of $G$
on $\BZ$ is regular. 

\subsubsection{Topology of $\BZ$}
\label{topologysect}
As a first step, we must determine the proper topology
of the Brillouin zone. 
According to the general theory~\cite[Corollary.~10.7]{Gallier-25-I},
the set $\widehat{N}$ of characters of a commutative locally compact group $N$ 
(i.e. the \emph{Pontryagin dual} of $N$) is a commutative locally compact group
for the pointwise product of characters
(i.e. $(\chi\cdot\chi')(g)=\chi(g)\chi'(g)$ for every $g$ in $N$)
where $\widehat{N}$ is equipped with the compact-open topology
(i.e. the topology of uniform convergence on compact sets).
For countable discrete groups, this means that
a sequence $\{\chi_n\}$ of characters of $N$ converges 
to the character $\chi$ iff for any $g$ in $G$
and any $\epsilon>0$, there is an integer $n_0$ such that
$|\chi_n(g)-\chi(g)| < \epsilon$ for all $n\ge n_0$. 
For the translation group $T$, this condition becomes
$|e^{-\imath \bfk_n\cdot\bft}-e^{-\imath \bfk\cdot\bft}| < \epsilon$
for all $n\ge n_0$. 
By using 
$|e^{-\imath \bfk_n\cdot\bft}-e^{-\imath \bfk\cdot\bft}|=|e^{-\imath (\bfk_n-\bfk)\cdot\bft}-1|$
and $2|\alpha|/\pi\le |e^{-\imath \alpha}-1|\le |\alpha|$ for $|\alpha|\le\pi$,
we see that, if $\bfk$ is in the interior of $\BZ$,
the compact-open topology on $\widehat{T}$ is equivalent to
the metric topology on $\BZ$, where the sequence
$\{\bfk_n\}$ converges to $\bfk$ iff, for every $\epsilon'>0$, there
is an integer $n_0$ such that $||\bfk_n-\bfk||<\epsilon'$ whenever $n>n_0$.
To be able to encompass points $\bfk$ on the surface of $\BZ$, we must take
the periodic property of $\BZ$ into account and
replace the distance $||\bfk-\bfk'||$ by
the distance $d(\bfk,\bfk')=\min_{\bfkm\in\Lambda^*}||\bfk-\bfk'+\bfkm||$.

\subsubsection{Countably-separated orbits}
Having identified the proper topology on $\BZ$, we
use the fact that $\BZ$ has
a countable family of open sets~\cite{Bourbaki-Topo-II} that we
build now.
Let $\{\bfk_n\}$ be a countable set of points of $\BZ$
which is dense in $\BZ$. For example, this can be
the set of points of $\BZ$ whose coordinates are rational numbers. 
We define the balls $U_{n,m}=\{\bfk\in \BZ; d(\bfk,\bfk_n)<1/m\}$,
where $d$ is the distance defined in the previous section.
These balls are a countable base of the topology of $\BZ$:
for any $\bfk$ in $\BZ$ and any open set $U$ of $\BZ$ containing 
$\bfk$, there is a ball $U_{n,m}$ such that
$\bfk\in U_{n,m}\subset U$. 

Let $g_1,\dots,g_q$ be representatives of the cosets of $T$ in $G$,
where $q={[}G{:}T{]}$ and let
$E_{n,m}=\{g_i\triangleright \bfk;  i=1,\dots,q\text{ and } \bfk\in U_{n,m}\}$,
where we recall that $g\triangleright\bfk=R\bfk$ with
$R$ the rotation part of $g$.
Since the action of $g$ on $\bfk$ does not involve the translation part of $g$,
we see that the orbit of the action of $G$ on $U_{n,m}$ is
$E_{n,m}$. In other words 
\begin{eqnarray*}
G\triangleright U_{n,m} &=& 
\{g\triangleright \bfk;  g\in G\,\,\text{and}\,\, \bfk\in U_{n,m}\} = E_{n,m}.
\end{eqnarray*}
The set $E_{n,m}$ is $G$-invariant because it is the orbit of $U_{n,m}$.
It is also a Borel set. Indeed, for every $i=1,\dots,q$,
the set
$E_i=\{g_i\triangleright \bfk; \bfk\in U_{n,m}\}$
is a Borel set, since it is the image of the Borel set $U_{n,m}$
under the homomorphism $\bfk\mapsto g_i\triangleright\bfk$
and $E_{n,m}=\bigcup_{1}^q E_i$ is a finite union of Borel sets.

We prove now that the collection $E_{n,m}$
separates the different orbits of $\BZ$. 
Let $\bfk$ and $\bfk'$ be two points of $\BZ$ such that
$G\triangleright\bfk\not=G\triangleright\bfk'$. Since orbits
are either identical or disjoint, this means that 
$g_i\triangleright\bfk\not=g_j\triangleright\bfk'$ for every $i$ and $j$
in $\{1,\dots,q\}$ and there is a minimal distance 
$\epsilon>0$ between the set 
$\{g_i\triangleright \bfk;  i=1,\dots,q\}$
and the set $\{g_i\triangleright \bfk';  i=1,\dots,q\}$.
Of course, we allow $g_i\triangleright \bfk=g_j\triangleright \bfk$
for some $i\not=j$.
By choosing $m>2/\epsilon$ and 
$n$ such that $d(\bfk,\bfk_n) < \epsilon/2$ and
$\bfk\in U_{n,m}$, we see that
$(G\triangleright\bfk')\cap E_{n,m}=\emptyset$
and $E_{n,m}$ separates the orbit of $\bfk$
from the orbit of $\bfk'$. 

To conclude that we have indeed condition (i),
we use the fact that any orbit $\calO$
is included in the intersection of the $E_{n,m}$'s containing
$\calO$. Then, for any orbit
$\calO'=G\triangleright\bfk'$ in this intersection,
we just proved that if $\calO'$ is not included
in $\calO$, then there is a $E_{n,m}$ that
separates it form $\calO$. This completes the
proof of condition (i).

\subsubsection{Irreducible Brillouin zone}

A first consequence of condition (i) is that
we can define an
irreducible Brillouin zone $\BZ/G=\BZ/G_0$. 
Indeed, it follows from (i) that 
we can find a set of representatives for $\BZ/G$
which is a Borel subset of $\BZ$.

We saw that, for $g= \cgop{R}{\bft}$ in $G$,
$g\triangleright \chi_{\bfk}=\chi_{R\bfk}$,
so that only the rotation part $R$ of $g$ 
acts on $\chi_\bfk$. Since $R$ runs over $G_0$
when $g$ runs over $G$, the quotient of the Brillouin
zone by $G$ is the irreducible Brillouin zone $\BZ/G_0$,
which is a convex subset of $\BZ$ such that
$\BZ$ is the union of $R\triangleright (\BZ/G_0)$
where $R$ runs over $G_0$ and the volume
of $\BZ/G_0$ is the volume of $\BZ$
divided by the number of elements of $G_0$.
In mathematical parlance,  irreducible Brillouin zones
are called \emph{orbifolds} and their construction is
gently described in Ref.~\cite{Zhilinskii}.
Irreducible Brillouin zones are widely used in
solid-state physics because they reduce the
computing time of band-structure calculations 
(see Refs.~\cite{Setyawan-10,Jorgensen-22}
for nice pictures of irreducible Brillouin zones). 

If regularity condition $(i)$ were not satisfied,
$\BZ/G$ would not be a measurable subset of $\BZ$.

\subsubsection{Proof of condition (ii)}
To show that condition (ii) holds for crystallographic groups,
we first recall that the quotient $G/H$
of a group $G$ by its subgroup $H$,
is the set of the left cosets of $H$ in $G$.

The translation group $T$ is clearly a subgroup of $G(\bfk)$.
Since the index $[G{:}T]$ of $T$ in $G$ is finite,
the index $[G(\bfk){:}T]$ is also finite and it divides $[G{:}T]$.
If we denote by $p=[G{:}G(\bfk)]$ the index of $H=G(\bfk)$ in $G$,
the quotient of $G$ by $H$ is $G/H=\{g_1\cdot H,\dots,g_p\cdot H\}$
and we can define a map $\varphi: G/H\to\calO_\bfk$ by
$\varphi(g_i\cdot H)=g_i\triangleright \bfk=R_i\bfk$, where
$R_i$ is the rotation part of $g_i$. 
Define the action of $G$ on $G/H$
by $g\triangleright (g_i \cdot H)=g_j \cdot H$,
where $g_j$ is the representative of $g\cdot g_i$
(i.e. there is a $h\in H$ such that $g\cdot g_i=g_j\cdot h$).
Then we see that the second condition is satisfied since 
$\varphi$ is a bijection between finite sets.


We are now ready to reap the fruits of our efforts.

\subsection{Mackey's imprimitivity theorems}
\label{imprimitivity-sect}

Mackey's work on the irreducible representations
of certain locally compact groups is a true masterpiece of mathematics~\cite{Varadarajan-08}.

Although the concept of imprimitivity is probably unknown to
most physicists, it has a very ancient and distinguished origin since
it can be traced back to Evariste Galois in the 1830'
as \emph{groupe non primitif}~\cite{Neumann-06,Brechenmacher-11}.
The concept was then more clearly defined
by Camille Jordan~\cite[p.~34]{Jordan}. It was translated into
german  by Dyck in 1883~\cite{Dyck-83} either as \emph{Imprimitivit{\"a}t}~\cite[p.~83]{Dyck-83}
or \emph{Nichtprimitivit{\"a}t}~\cite[p.~86]{Dyck-83}
and stabilized as \emph{imprimitive Gruppe} by Frobenius in 1895~\cite{Frobenius-95}.
Burnside translated it into \emph{imprimitive group} in 1897~\cite[p.~193]{Burnside}.
The relevance of these concepts for locally compact groups was 
stressed by Mackey~\cite{Mackey-49}.


We now present a theorem which is equivalent to step (vi) showing that
all irreducible representations of $G$ can be obtained by the construction.
We first recall the definition of a representation $\sigma$ of $G$
induced from a representation $\rho$ of $G(\bfk)$.
Let $g_j$ with $j=1,\dots,|G/G(\bfk)|$  be a complete set of  representatives
of the cosets of $G(\bfk)$ in $G$ and let
\begin{eqnarray}
\sigma(g)_{ir,js} &=& 
\sum_{h\in G(\bfk)} \rho(h)_{rs} \delta_{h,g_igg_j^{-1}}.
\label{inducedrep}
\end{eqnarray}
Then, $\sigma$ is a representation of $G$ called the representation of $G$
induced from the representation $\rho$ of $G(\bfk)$. The following
theorems describe the nature of these induced representations.

The first theorem~\cite[Thms.~7.4]{Gallier-25}
states that, since the action of $G$ on $\BZ$ is regular,
any unitary irreducible representation of $G$
can be obtained by induction from an irreducible
representation of a little group $G(\bfk)$.

Let $G$ be a crystallographic group and $T$ its subgroup of translations.
For every unitary irreducible representation $(\calH_\sigma,\sigma)$ of
$G$, there is unique orbit $\calO$ and, for any $\bfk$
satisfying $\calO_\bfk=\calO$, there is a unitary irreducible 
representation $(\calH_\rho,\rho)$ of $G(\bfk)$ 
such that $\sigma$ is equivalent to 
the representation of $G$ induced from
the representation $\rho$. 

As in the case of finite space groups~\cite{Kim},
we see that the irreducible representations of $G$
come from the \emph{small representations} $(\calH_\rho,\rho)$
of some little group $G(\bfk)$, which are
representations such that, for any $t\in T$, 
$\rho(t)=\chi_\bfk(t)\id_{\calH_\rho}$.

We just showed that each irreducible representation of $G$ can be obtained from an
irreducible representation of some $G(\bfk)$. Now we deal with
step (v) by proving that each representation induced from a small representation of 
a little group is irreducible. This is the 
second fundamental theorem~\cite[Thms.~7.5]{Gallier-25},
which holds again because the action of $G$ on $\BZ$ is regular:

Let $G$ be a crystallographic group and $T$ its subgroup of translations.
For any $\bfk\in\BZ$ and for any small representation 
$(\calH_\rho,\rho)$ of $G(\bfk)$, the representation 
of $G$ induced from $\rho$ is an irreducible 
representation of $G$.

\subsubsection{Constructing the irreps of the little groups}
We used a rather heavy mathematical machinery to prove
that all the irreps of a crystallographic group $G$ can be obtained form
irreps of $G(\bfk)$ for some $\bfk$.
This may seem to be a small progress because 
$G(\bfk)$ is also a crystallographic group with an infinite number of irreps. 
Moreover, $G(\bfk)=G$ when $\bfk=0$.
However the crucial point is that we need only the small 
representations of $G(\bfk)$ and there are
only a \emph{finite} number of them.

Since $T$ is an invariant subgroup of $G(\bfk)$,
we can obtain representations of $G(\bfk)$ by induction
from representations $\chi_\bfk$ of $T$.
We must show that these induced representations are
small representations and that all small representations
can be obtained by such an induction.
In the finite group case, this is done by using
the Frobenius reciprocity theorem~\cite[Section~14.4]{Kim}.

In locally compact groups the Frobenius reciprocity theorem is not always valid,
but it turns out that it is valid for countable discrete groups such as 
crystallographic groups~\cite{Mackey-52}.
Therefore, the proof of the finite group case can
be followed (but see Ref.~\cite{Mackey-58-2} for a proof in the context of
locally compact groups) and the small representations of
$G(\bfk)$ are built in terms of projective
representations of the point group $G_0(\bfk)=G(\bfk)/T$, which is called
the \emph{cogroup} of $\bfk$.
Note that $G_0(\bfk)$ is a subgroup of the point group $G_0$ of $G$
because $T$ is an invariant subgroup of $G(\bfk)$.

The cogroup $G_0(\bfk)$ is the set of elements  $R$ of $G_0$ 
such that $R\bfk=\bfk+\bfkn$ for some $\bfkn\in\Lambda^*$.
Let $(V,\rho)$ be a  small representation of $G(\bfk)$.
We first notice that $\rho$ is entirely determined on $T$ because,
by definition of small representations, in any basis of $V$
\begin{eqnarray*}
\rho\big(\cgop{\un}{\bftm}\big) &=&  \chi_\bfk\big(\cgop{\un}{\bftm}\big)\id_d = e^{-\imath\bfk\cdot\bftm} \id_d,
\end{eqnarray*}
where $d=\dim V$. 

On the other hand, we know that all elements of $G$ can
be written  $\cgop{R}{\bft_R+\bftm}$ for some $\bftm\in\Lambda$. 
From  $\cgop{R}{\bft_R+\bftm}=\cgop{\un}{\bftm}\cdot\cgop{R}{\bft_R}$,
we can write
\begin{eqnarray}
\rho\big(\cgop{R}{\bft_R+\bftm}\big) &=&  \rho\big(\cgop{\un}{\bftm}\big) \rho\big(\cgop{R}{\bft_R}\big)
= e^{-\imath\bfk\cdot\bftm}\rho\big(\cgop{R}{\bft_R}\big).
\label{rhooftau}
\end{eqnarray}
Therefore, $\rho$ is entirely determined if we know
$\rho\big(\cgop{R}{\bft_R}\big)$ for each $R\in G_0(\bfk)$.
Choose a collection of representatives $g_1=\cgop{R_1}{\bft_1},\dots, g_r=\cgop{R_r}{\bft_r}$,
with  $r=|G_0(\bfk)|$ of left cosets of $T$ in $G(\bfk)$,
where $R_1,\dots,R_r$ are the elements of the group $G_0(\bfk)$.
Every  $\cgop{R}{\bft}$ of $G(\bfk)$ can be written $\cgop{R_i}{\bft_i}\cdot\cgop{\un}{\bftm}$ for a unique $i$
and a unique $\bftm\in\Lambda$.
To determine $\rho(g_i)$ we compute
\begin{eqnarray*}
g_i\cdot g_j &=& \cgop{R_iR_j}{R_i\bft_j+\bft_i}. 
\end{eqnarray*}
Let $R_iR_j=R_k$ in $G_0(\bfk)$. There is a unique 
$\bftm\in\Lambda$ such that 
\begin{eqnarray*}
g_i\cdot g_j &=& \cgop{R_iR_j}{R_i\bft_j+\bft_i} =\cgop{R_k}{\bft_k+\bftm}=
\cgop{\un}{\bftm}\cdot g_k.
\end{eqnarray*}
This has two consequences: firstly
$R_i\bft_j+\bft_i-\bft_k=\bftm\in\Lambda$ and secondly
\begin{eqnarray}
\rho(g_i\cdot g_j) &=& \rho\big(\cgop{\un}{\bftm}\cdot g_k\big)=
\rho\big(\cgop{\un}{\bftm}\big)\rho(g_k)=e^{-\imath\bfk\cdot\bftm}\rho(g_k),
\label{taugigj}
\end{eqnarray}
Since the point group $G_0(\bfk)$ is the quotient of $G(\bfk)$ by the
invariant subgroup $T$, we can consider that 
$G_0(\bfk)=\{R_1,\dots,R_r\}$ and we define the map
$\tau: G_0(\bfk) \to M_d(\bbC)$ by $\tau(R_i)=\rho\big(\cgop{R_i}{\bft_i}\big)$.
According to Eq.~\eqref{taugigj} we have
\begin{eqnarray*}
\tau(R_i)\tau(R_j) &=& e^{-\imath\bfk\cdot(R_i\bft_j+\bft_i-\bft_k)}\tau(R_i R_j).
\end{eqnarray*}
It can be shown that the phase factor satisfies a cocycle condition~\cite{Hiller-86}
and that $\tau$ is a projective representation~\cite[Chapter~12]{Kim} of $G_0(\bfk)$.
The progress is that $G_0(\bfk)$ is a finite group and finite groups
have only a finite number of irreducible projective representations
up to equivalence~\cite[p.~570]{Karpilovsky}.

\subsubsection{A happy conclusion}
We gave an example of the fact that the action of $\bbZ$ on 
the circle can be ergodic if an irrational number is involved.
By construction, the wavevectors of a finite toric crystal
are fractions of the reciprocal basis vectors. It could have been
feared that the action of a crystallographic group on
a wavevector involving irrational numbers might have
been ergodic. We showed that this is not the case.

By using the regularity of the action of a crystallographic group $G$ on 
the Brillouin zone $\BZ$ and 
the Mackey machine, we showed that, for every wavevector $\bfk\in\BZ$,
the construction of the irreducible representations of $G$ corresponding to
$\bfk$ proceeds from the finite projective representations of the finite
group $G_0(\bfk)$ as for the case of a finite toric crystal (for which
the components of $\bfk$ are rational numbers).
We also showed that all the irreducible representations of $G$
are obtained by this construction. 
Therefore, the happy conclusion of our work is that
the worst behaviors of infinite groups do not take
place in crystallographic groups and the construction of their
irreducible representations is similar to that of
finite toric crystals.

\section{Irreps of infinite magnetic groups}

A magnetic group (also called a $C_2$-graded group in
the mathematical literature~\cite{Rumynin-21}) is a
group $G$ that can be written as 
$G=G_+ \cup G_-$, where $G_+$ is a subgroup of $G$
of index 2. 
What makes it magnetic is that $G$ is not represented by
unitary representations but by \emph{corepresentations},
which means that elements of  $G_+$, denoted by $u$,
are represented by unitary operators while
elements of  $G_-$, denoted by $a$, are represented by antiunitary operators.  

Magnetic groups can also be used to describe non-magnetic properties
such as crystal twinnings~\cite{Curien-58,Hahn-Klapper}.

Magnetic crystallographic groups 
in general dimensions received much less attention than their non-magnetic counterparts
and they have been enumerated only in dimension $n=$1, 2, 3 and 4~\cite{Souvignier-06}
(see Table~\ref{table-groupe-magnetique}).

\begin{table}
	\caption{Number of magnetic point and space groups for 
		dimensions 1 to 4~\cite{Souvignier-06}. Magnetic groups are
		considered to be equivalent if they are related by an isometry
		conserving the orientation. Magnetic point groups are
		of type I if they do not involve time-reversal, of type
		II if time-reversal is an element of the group,
		of type IV if time-reversal multiplies a non-zero translation
		and of type III otherwise~\cite{Cracknell}.
	}.
	\label{table-groupe-magnetique}
	\begin{center}
		\begin{tabular}{|c|c|c|c|c|} \hline
			Dimension & 1 & 2 & 3 & 4 \\
			\hline 
			Magnetic point groups        & 5 & 31 & 122 & 1202 \\
			Magnetic point groups (I/II) & 2 & 10 & 32 & 271 \\
			Magnetic point groups (III)  & 1 & 11 & 58 & 660 \\
			Magnetic space groups        & 7 & 80 & 1651 & 62227 \\
			Magnetic space groups (I/II) & 2 & 17 & 230 & 4894 \\
			Magnetic space groups (III)  & 1 & 26 & 674 & 30497 \\
			Magnetic space groups (IV)  & 2 & 20 & 517 & 21942 \\
			\hline
		\end{tabular}
	\end{center}
\end{table}

The particularly important case of the 1651 three-dimensional magnetic groups
was classified in 1957 by Nikolai Vasil'evich Belov, Nina Nikolaevna Neronova and
Tamara Serafimovna Smirnova~\cite{Belov-57}
and in 1963 by Rosalia Guccione~\cite{Guccione-63,Opechowski-65}
(see Ref.~\onlinecite{Grimmer-08} for a critical comparison of the two approaches).
Litvin described these groups in extensive (12086 pages long) tables~\cite{Litvin-14}.
Two-dimensional magnetic groups
were classified in 1935 by Henry John Woods 
to describe black and white textile patterns~\cite{Woods-35-III}.

From the mathematical point of view,  there are very few works on the corepresentation of 
locally compact groups by unitary and antiunitary operators~\cite{Neeb-17}.

The theory of corepresentations of \emph{finite} magnetic groups was described
by Wigner in his book on group theory~\cite[p.~336]{Wigner}. 
We derive it now for infinite crystallographic groups. 
We first describe corepresentations, then we show how 
corepresentations can be expressed in terms of
unitary representations of $G_+$.

\subsection{Corepresentations of magnetic groups}
We recall that a magnetic group $G$ can be written as 
$G=G_+ \cup G_-$, where $G_+$ is a subgroup of $G$
of index 2, and thus an invariant subgroup of $G$
(the proof is usually done for finite groups~\cite[Example~7.6]{Humphreys} 
but it extends without change to infinite groups.
Moreover, $G_+$ is a crystallographic group
because every subgroup of finite index
of a crystallographic group in $\bbR^n$ is a crystallographic group
in $\bbR^n$~\cite[Thm.~4]{Farkas-81}.

If $T$ (resp. $T_+$) denotes the translation group of $G$ (resp. $G_+$)
and $G_0$ (resp. $G_{+0}$) its point group, then we have
only two possibilites: either $T_+=T$ and $G_{+0}$ is a subgroup of
$G_0$ of index 2 ($G_+$ is then said to be a 
\emph{translationgleich} subgroup of $G$), or $G_{+0}=G_0$ and $T_+$ is a subgroup of $T$
of index 2 ($G_+$ is then said to be a 
\emph{klassengleich} subgroup of $G$)~\cite[Thm.~17.1]{Opechowski}.

Since $ G_+$ is a subgroup of $G$ of index 2, 
$G_-$ is a left coset $G_-=r G_+$ for any $r\in G_-$.
The quotient  $G/ G_+$ is a group containing two elements: the identity
$e$ representing the elements of $ G_+$ and the element $\xi$ representing the
elements of $G_-$. There is a homomorphism from $G$ to 
$G/ G_+$~\cite[p.~12]{LudwigFalter}. In particular, since $G/ G_+$ is a group,
the inverse of $\xi$ can only be $\xi$ and $\xi^2=1$. As a consequence
of the homomorphism, $(r u_1)(r u_2)\in  G_+$ for any $u_1$ and $u_2$ in $G_+$. In particular $r^2\in G_+$. 

Let us define antiunitary operators.
An operator $O:\calH\to\calH$ acting on a Hilbert space $\calH$ is antilinear
if $O(v_1+v_2)=O(v_1) + O(v_2)$ and 
$O(\lambda v)=\overline{\lambda} O(v)$,
for every $v$, $v_1$ and $v_2$ in $\calH$ and $\lambda$ in $\bbC$, where $\overline{\lambda}$
is the complex conjugate of $\lambda$. 
The adjoint $O^\dagger$ of the antilinear operator $O$  is defined by  
$\langle v_1, Ov_2\rangle=\langle v_2, O^\dagger v_1\rangle$.
Then, an operator $O$ is antiunitary if it is antilinear
and $O^\dagger O=O O^\dagger =\id$.

A corepresentation of $(\calH,\sigma)$ of $G$ 
	is a homomorphism $\sigma$ of $G$ into the group  of 
	all unitary or antiunitary operators on $\calH$ such that 
	$\sigma(g)$ is unitary for every $g\in G_+$ and antiunitary for every $g\in G_-$.
	The corepresentation $\sigma$ is said to be reducible
	if $\calH$ and $\{0\}$  are the only closed $\sigma(G)$-invariant subspaces of 
	$\calH$.
	
	\subsection{Corepresentation of $G$ from representation of $G_+$}
	We show now that an irreducible corepresentation of $G$ can be expressed in terms of
	a representation of the subgroup $G_+$. 
	Let $(\calH,\sigma)$
	be an irreducible corepresentation of $G$.
	The restriction $\sigma|_{G_+}$ of $\sigma$ to $G_+$ is
	a unitary representation of the crystallographic group $G_+$.
	If  we could find a  $\sigma(G_+)$-invariant subspace $V$ of  $\calH$
	such that the representation  of the crystallographic group
	$G_+$ defined on $V$ is irreducible, we could proceed
	with the  analysis of $\sigma.$ However, for an infinite group $G$, a subgroup 
	$H$ and an irreducible representation $(\calH,\sigma)$ of $G,$ 
	it may happen that such a subspace $V$ does not exist. 
	
	For instance, let $G$ be the countable discrete group   $\left\{g_{ab}=\begin{pmatrix} a&b\\0&1\end{pmatrix}: a, b\in \mathbb{Q}, a\neq 0  \right\}$
	and fix a non-trivial unitary character $\delta$ of $\mathbb{Q};$  the representation $\sigma$ of $G$ on $\ell^2(\mathbb{Q}\setminus\{0\})$
	defined by $\sigma(g_{ab})f (x)= \delta(bx) f(ax)$ is irreducible (see~\cite[Rem.3.C.5]{Bekka-21}).
	The subgroup $H=\left\{g_{a0}: a\in\mathbb{Q}\setminus\{0\}\right\}$
	is a  subgroup of $G$ but there is no subspace of $\ell^2(\mathbb{Q}\setminus\{0\})$ which is invariant and irreducible under 
	$\sigma(H)$.
	Indeed, as $H$ is commutative, such a subspace $V$ would have to be one-dimensional 
	and there would be a non-zero function $f$ and a unitary character $\chi$ of $H$ 
	such that $\sigma(g_{a0})f=\chi(g_{a0})f$, or $f(ax)=\chi(g_{a0})f(x)$.
	As a consequence, the function $|f|$ would satisfy $|f(ax)|=|f(x)|$  for every $a\in\mathbb{Q}\setminus\{0\}$
	but the zero function is the only dilation invariant function in $\ell^2(\mathbb{Q}\setminus\{0\}).$
		
	Coming back to our irreducible corepresentation $(\calH,\sigma)$ of the magnetic $G$,
	it has been shown~\cite[Theorem 2.11.c]{Neeb-17} (see also Wigner~\cite{Wigner} for the case
	where $\calH$ is finite dimensional) that 
	either $\sigma|_{G_+}$ is irreducible or $\sigma|_{G_+}$ is a direct sum of two 
	irreps of $G_+.$
	As we know that the irreps of a crystallographic group are finite dimensional,
	we see that there is a finite dimensional  $\sigma(G_+)$-invariant subspace $V$ 
	which is irreducible under $G_+$.
	
	Let $\psi_1,\dots,\psi_d$ be a basis of $V$.
	The action of $G_+$ on $V$ 
	is written
	\begin{eqnarray*}
		u\triangleright \psi_i &=& \sum_j \rho(u)_{ji}\psi_j.
	\end{eqnarray*}
	For an arbitrary $r$ in $G_-$, set $\psi'_i=r\triangleright \psi_i$. 
	The action of an element $u$ of $G_+$ on $\psi'_i$ is 
	\begin{eqnarray*}
		u\triangleright \psi'_i &=& (ur)\triangleright\psi_i=r(r^{-1}ur)\triangleright\psi_i.
	\end{eqnarray*}
	Since $r^{-1}ur\in  G_+$ it can be represented by $\rho(r^{-1}ur)$ and
	\begin{eqnarray*}
		u\triangleright \psi'_i &=& r\triangleright \sum_j \rho(r^{-1}ur)_{ji}\psi_j
		= \sum_j \overline{\rho(r^{-1}ur)_{ji}}\,\,r\triangleright\psi_j
		\\&=&
		\sum_j \overline{\rho(r^{-1}ur)_{ji}}\psi'_j,
	\end{eqnarray*}
	because the action of $r$ is antilinear.
	Denote $W$ the vector space generated
	by $\psi'_1,\dots,\psi'_d$ and
	$\tau(u)=\overline{\rho(r^{-1}ur)}$.
	It is easy to see that $(W,\tau)$ 
	is a unitary representation of $G_+$
	and $\tau$ is irreducible, because
	otherwise $\rho$ would also not be irreducible.

From now on, we can follow without change the argument
given by Wigner~\cite{Wigner,Wigner-60,Loring-25}
to show that, either $W=V$ or $W\cap V=\{0\}$
and, using the fact that Schur's lemma 
holds for general topological groups~\cite[Prop.1.A.11]{Bekka-21} and in particular
for infinite crystallographic groups, we get 
the classification of corepresentations of magnetic groups.

\subsection{Classification of corepresentations}
Let $\sigma$ be an irreducible corepresentation of a 
magnetic crystallographic group $G=G_+ \cup G_-$ and 
let $\rho$ be the smallest dimensional irreducible
representation of $G_+$ in the restriction $\sigma|_{G_+}$.
We classify $\sigma$ by comparing $\rho$
with the irreducible representation $\tau$ of $G_+$ defined by
$\tau(u)=\overline{\rho(r^{-1}ur)}$,
where $r$ is an arbitrary element of $G_-$.

There are three cases
\begin{enumerate}
	\item If $\rho$ and $\tau$ are equivalent, then there is 
	a unitary matrix $M$ such that $\tau(u)=M^{-1} \rho(u) M$.
	Wigner showed that $M\overline{M}=\pm\rho(r^2)$.
	\begin{enumerate}
		\item (first type for~\cite[p.~343]{Wigner}, type (a) for~\cite[p.~616]{Bradley-10}): If $M\overline{M}=+\rho(r^2)$,  
		then the irreducible corepresention $\sigma$ is equivalent to 
		\begin{eqnarray*}
			\sigma'(u) &=&\rho(u),
			\\
			\sigma'(a) &=& \rho(ar^{-1}) M,
		\end{eqnarray*}
		its representation space is $V=W$ and the dimension of $\sigma$ is the same as the dimension of $\rho$.
		\item (second type for~\cite[p.~344]{Wigner}, type (b) for~\cite[p.~616]{Bradley-10}): If $M\overline{M}=-\rho(r^2)$, the 
		irreducible corepresentation $\sigma$ is equivalent to 
		\begin{eqnarray*}
			\sigma'(u) &=& \ppmatrix{\rho(u)}{0}{0}{\rho(u)},
			\label{canonicalbu}\\
			\sigma'(a) &=&   
			\ppmatrix{0}{\pm \rho(ar^{-1})M}{\rho(ar^{-1})M}{0},
			\label{canonicalba}
		\end{eqnarray*}
		its representation space is $V\oplus W$ and its dimension is twice the dimension of $\rho$.
	\end{enumerate}
	\item (third type for~\cite[p.~343]{Wigner}, type (c) for~\cite[p.~616]{Bradley-10}): If $\rho$ and $\tau$ are not equivalent, then the corepresentation $\sigma$ is equivalent to
	\begin{eqnarray*}
		\sigma'(u) &=& 
		\left( \begin{array}{cc}
			\rho(u) & 0 \\ 
			0 & \overline{\rho(r^{-1}ur)}
		\end{array}\right),
		\\
		\sigma'(a) &= &
		\left( \begin{array}{cc}
			0 & \rho(ar) \\
			\overline{\rho(r^{-1}a)} & 0
		\end{array}\right),
		\label{canonical}
	\end{eqnarray*}
	its representation space is $V\oplus W$ and its dimension is twice the dimension
	of $\rho$.
\end{enumerate}
An important conclusion is that all irreducible corepresentations of
magnetic crystallographic groups are finite dimensional
and are classified as in the finite toric crystal case. 
The permutation group reduction of tensor powers of corepresentations
	was discussed by Gard and Backhouse~\cite{Gard-75}.

\section{Induced representation theorems for crystallographic groups}
Induced representations are a standard tool of quantum physics
and books are essentially devoted to their enumeration~\cite{Kovalev}
and to their applications in the elecron theory of solids, for
the description of defects in crystals and 
second-order phase transitions,
as well as in phonon spectroscopy and 
magnetism~\cite{Altmann-77,Evarestov-97,Evarestov-12}.

As a simple example of induced representations,
consider the computation of molecular orbitals
of methane~\cite{Yarzhemsky-19,Yarzhemsky-25}.
The methane molecule (CH$_4$) has the symmetry group $T_d$
while each hydrogen atom has only the symmetry subgroup $C_{3v}$.
The irreducible representation describing the atomic orbitals of H 
can generate orbitals of the whole molecule
by induction of an irreducible representation of 
$C_{3v}$ to a representation of $T_d$.
Similarly, the irreducible representations of $C_{3v}$ describing the 
vibration of a hydrogen atom can be used to induce
a representation of $T_d$ describing the vibration of all the hydrogen atoms in
the methane molecule.

In this section we discuss several operations that can be
carried out with induced representations : restriction of an induced
representation to a subgroup, tensor product of induced representations,
symmetric and antisymmetric squares of induced representations.

As a word of caution it must be stressed that,
although induced representations of locally compact groups
are well defined, the induced representation of 
more general infinite groups from a unitary representation 
of a subgroup does not always exist, and when it does
it can be not uniquely defined~\cite{Kosyak-14,Kosyak-18}.

\subsection{Double cosets}
In section~\ref{imprimitivity-sect} we considered crystallographic groups 
$G$ and subgroups $H$ of $G$ of finite index $[G{:}H]$.
In the present section we consider general sugroups $H$ 
that can be finite, for example of the finite symmetry group (i.e. stabilizer) of a particular
site of the crystal, or infinite, for example
the infinite subgroup of $H$ that stabilizes a plane in
$\Lambda$ (used in surface physics).
In both cases, the index of the subgroup is infinite 
and we must deal with an infinite
set of coset representatives $\{g_j\}_{j\in J}=G/H$ where $J$ is 
a countable set. The choice of $g_j$ in each coset must
be explicit for practical applications
(and to avoid the Axiom of choice).
We still have the decomposition
\begin{eqnarray*}
G &=& \bigsqcup_{j\in J} g_j H.
\end{eqnarray*}

Since this section relies on double cosets, we first define this concept. 
Let $G$ be a group, $H$ and $K$ two subgroups of $G$. 
For each $g\in G$, the \emph{double coset} (or $(H,K)$-double coset) of $g$ is the 
set  $HgK=\{h g k \text{ for all } h \text{ in } H \text{ and } k \text{ in } K\}$.
Two elements $g$ and $g'$ of $G$ belong to the same
double coset (or are equivalent for the double-coset relation)
if there is a $h$ in $H$ and a $k$ in $K$ such that
$g'=hgk$. Two double cosets are either identical or disjoint.
The set of all double cosets is denoted by 
$\{g_j\}_{j\in J}=H\backslash G/K$ where $J$ is 
a countable set and $g_j$ is a \emph{representative} of the double coset
$H g_j K$. 
The group $G$ is the disjoint union of its double cosets
\begin{eqnarray*}
	G &=& \bigsqcup_{j\in J} Hg_j K.
\end{eqnarray*}
Again, for practical applications, it is
useful to explicitly determine the representatives
$g_j$, as done for halite in 
Sect.~\ref{halitedoublecosetsect}.

Double cosets are a bit like cosets but do not share all their
properties: different double cosets can have different numbers
of elements, which do not necessarily divide the order of $G$.

In this section, the subgroup $g_j K  g_j^{-1}$ is often used.
This is because there is an obvious bijection between
$Hg_jK$ and $Hg_j K  g_j^{-1}$, which is a product 
subgroup $H$ and subgroup $Hg_j K  g_j^{-1}$.
The structure of such product of groups can be seen
as the quotient of the product $H\times K$
by the common subgroup $K_j=g_j K  g_j^{-1}\cap H$. 
This provides an explicit description of double cosets
in terms of right-cosets of $H$ and left-cosets of $K$.
For example,  let 
$\{h_{p}\}_{p\in P}$ be a complete set of representatives of the 
right-coset decomposition of $K_j$ in $g_j K g_j^{-1}$:
\begin{eqnarray*}
	g_j K g_j^{-1} &=& \bigsqcup_{p\in P} K_j h_{p}. 
\end{eqnarray*}
Then, 
\begin{eqnarray*}
	H g_j K &=& \bigsqcup_{p\in P} H h_{p}g_j,
\end{eqnarray*}
is the partition of $H g_j K$ into
right cosets of $H$.

\subsection{Mackey's restriction formula}

Mackey's restriction theorem~\cite{Mackey-51} gives a precise solution to the following 
problem. Consider a group $G$, two subgroups 
$H$ and $K$ and a representation $(\calH_\rho,\rho)$ of $K$.
If $\sigma$ is the representation of $G$ induced from $\rho$,
what is the result of the restriction of $\sigma$ to the subgroup $H$?
For example, if we consider a space group $G$ and
an atom at (Wyckoff) positions $(x)$, we can build
an orbital of the crystal by induction from an orbital
of the atom. This is a classical problem of solid-state physics
that has been discussed from the group-theoretical point of view
since the early sixties~\cite{Levinson-61,Cloizeaux-63}
(see Ref.~\cite{Evarestov-97} for more details).
In general we start from an atomic orbital
which is an irrep of the symmetry point group (i.e. stabiliser) $K$
of position $(x)$ and induce a representation of $G$
by the usual formula. This infinite-dimensional representation 
can be expanded into irreducible representations of $G$, but
Mackey's restriction formula enables us to avoid this step. 
Now we wonder how this induced representation is seen from another
atom at (Wyckoff) positions $(y)$ with symmetry point group $H$.
In mathematical terms, we induce a representation $\sigma$ of $G$ from the representation
$\rho$ of $K$ and we compute the restriction of $\sigma$ to $H$.

\emph{Mackey's restriction formula} precisely describes
this situation for finite groups~\cite{Mackey-51} and
then for some locally compact groups~\cite{Mackey-52}.
Fortunately, crystallographic groups satisfy the conditions of
validity of Ref.~\cite{Mackey-52} because
two arbitrary subgroups $H$ and $K$ of a crystallographic group $G$ are \emph{discretely related}
in the sense of section~7 of Mackey's paper~\cite{Mackey-52} since
$G$, being a countable set,  is a countable union of
double cosets $H\backslash G/K$. 

To state Mackey's restriction formula,
let $\{g_j\}_{j\in J}$, where $J$ is a countable set,
be a complete set of representatives 
of the double cosets $H\backslash G/K$, so that 
$G=\cup_{j\in J} H g_j K$. Define the sugroups $K_j=g_j K g_j^{-1} \cap H$
of $H$. If $(\calH,\rho)$ is a representation of $K$, then
$\rho_j(k)=\rho(g_j^{-1}kg_j)$, for $k\in K_j$, is a representation of $K_j$.
If $\sigma$ is the representation induced from $\rho$ to $G$,
Mackey's restriction formula~\cite[Thm.~7.1]{Mackey-52} states that the restriction of $\sigma$ to $H$ is
\begin{eqnarray*}
(\ind_K^G\rho)|_H  &=&
\sigma|_H = \bigoplus_{j\in J} \ind_{K_j}^H\rho_j.
\end{eqnarray*} 

We describe more explicitly $\ind_{K_j}^H\rho_j$. Since $K_j$ is a subgroup
of $H$, we can write $H=\cup_i h_i K_j$, where $\{h_1,\dots,h_q\}$ is a complete
set of representatives of the cosets of $K_j$ in $H$. 
For $k$ an element of $K_j$, we know that $g_j^{-1}kg_j$ is an element
of $K$ and $\rho_j(k)=\rho(g_j^{-1}kg_j)$.
By definition~\eqref{inducedrep} of induced representations,
if $\tau_j=\ind_{K_j}^H\rho_j$, then
\begin{eqnarray*}
\big(\tau_j(h)\big)_{mr,ns} &=& 
\sum_{k\in K_j} \rho_j(k)_{rs} \delta_{k,h_m h h_n^{-1}}.
\end{eqnarray*}

Mackey's restriction formula was used in
concrete physical applications~\cite{Bradley-66,Evarestov-97}.

\subsection{The example of halite}
As an illustration of Mackey's restriction formula,
we consider the case of the mineral
halite (which is the common rock-salt NaCl).
Halite is a ionic crystal made of ions Na$^+$
and Cl$^-$. The 3s valence electron of sodium
(i.e. described by the fully symmetric representation of the point
symmetry group $O_h$ of the Na site)
can be used to induce an infinite dimensional
representation of the crystal group, as done by the Linear Combination
of Atomic Orbitals technique~\cite{Evarestov-12}.
To see how this state can hybridize with
the 3p hole of chlorine, we project this
representation onto the $T_{1u}$ 
irrep at the Cl site with symmetry $O_h$. 
Mackey's restriction formula enables us to
do that without computing the induced representation.

Another common physical problem is
to consider the vibrations of sodium atoms
(which are  represented by the $T_{1u}$ irrep of $O_h$),
to build an infinite-dimensional representation
of the crystallographic group (describing vibrations of the full crystal)
and to see how this state can hybridize with the vibrations at the chlorine site,
which are also represented by the 
$T_{1u}$ irrep of $O_h$.

\subsubsection{Halite structure}
The three-dimensional space group halite is
$Fm\bar{3}m$ (No.~225 in the Tables~\cite{Hahn}).
In an orthonormal reference frame
$\{\bfe_x,\bfe_y,\bfe_z\}$ where the
unit of length is $5.6404\times 10^{-10}~m$, 
the usual primitive basis of the primitive cell
is $\bfa_1=(\bfe_y+\bfe_z)/2$,
$\bfa_2=(\bfe_x+\bfe_z)/2$ and $\bfa_3=(\bfe_x+\bfe_y)/2$.
In that cell, 
Na atom is at $\bfx=(0,0,0)$, i.e. Wyckoff position
$(4a)$, and  
Cl atom at $\bfy=(1/2,1/2,1/2)$, i.e. Wyckoff position
$(4b)$. 
Geometrically, Na atoms sit at the vertices of
the parallelepiped generated by $\bfa_1$,
$\bfa_2$ and $\bfa_3$ and a Cl atom sits at 
its center, and this parallelepiped is repeated
in all directions to fill space. 

\subsubsection{Halite space group}
Space group $G=Fm\bar{3}m$ is symmorphic
(i.e. the semi-direct product of $G_0$ and $T$) and 
its elements are of the form
$\cgop{R}{\bftm}$, where $\bftm\in\Lambda$
and $R$ runs over the point
group $G_0$ denoted by
$m\bar{3}m$ (Hermann-Mauguin notation)
or $O_h$ (Schoenflies notation)
with 48 elements.

It is convenient to use the orthonormal reference frame
$(\bfe_x,\bfe_y,\bfe_z)$. The cube generated by these
three vectors is called the \emph{conventional cell}
and its volume is four times the volume of the primitive cell.
Na atoms sit at the vertices of this cube and at the center
of its faces, while Cl atoms sit at the center of the cube
and at the center of its vertices. 
In the crystallographic notation, the elements
$R$ of $G_0$ are denoted by its action on a
vector $\bfr=(x,y,z)$. 
For example, a rotation of $\pi/2$ around the $z$ axis
is denoted by $(y,-x,z)$ or the inversion by $(-x,-y,-z)$.
Therefore, each of the 48 operations of $G_0$
is represented by a permutation of some
$(\epsilon_1 x,\epsilon_2 y,\epsilon_3 z)$,
where $\epsilon_i=\pm1$.
As a consequence, the orbit of $\bfr$ under the action of $G_0$ is
determined by its element with coordinates $0\le x\le y \le z$ 
and the number of elements of this orbit
varies from 1 if $x=y=z=0$ (i.e. $\bfr=\bfx$) to 48 if $0<x<y<z$.
In particular, the orbit of $\bfy$ is made of the eight 
points $(\epsilon_1/2,\epsilon_2/2,\epsilon_3/2)$.
In the conventional cell, the translation vectors
$\bftm$ of $\Lambda$
are now
\begin{eqnarray*}
\bftm &=& m_1\bfa_1+m_2\bfa_2+m_3\bfa_3 = \frac{m_2+m_3}{2}\bfe_1 + \frac{m_1+m_3}{2}\bfe_2 + \frac{m_1+m_2}{2}\bfe_3
=r_1\bfe_1 + r_2\bfe_2 + r_3\bfe_3.
\end{eqnarray*}
Note that $r_1$, $r_2$ and $r_3$ are either all integers or
one integers and two half-odd-integers.

\subsubsection{Symmetry group of Na and Cl}

The symmetry group of a site $\bfr$ is its stabilizer, i.e.  the set of
operations $\cgop{R}{\bftm}$ such that
\begin{eqnarray}
\cgop{R}{\bftm}\triangleright \bfr &=& 
R\bfr+\bftm=\bfr.
\label{Rbfr}
\end{eqnarray}

If $\bfr=\bfx=0$, this gives us the 
$\bftm=0$ for all $R$ in $G_0$.
Therefore the symmetry group of 
Na is $K=\{\cgop{R_q}{0}, R_q\in G_0\}$.
If $\bfr=\bfy=(1/2,1/2,1/2)$ Eq.~\eqref{Rbfr} becomes
\begin{eqnarray*}
R(1/2,1/2,1/2)+\bftm=(1/2,1/2,1/2).
\end{eqnarray*}
If $R_p$ in $G_0$ is represented by
$(\epsilon_1 r_1,\epsilon_2 r_2,\epsilon_3 r_3)$,
where $(r_1,r_2,r_3)$ is a permutation of $(x,y,z)$,
then the corresponding $\bftm$ is 
\begin{eqnarray*}
{\bftm}_p &=& \sum_{i=1}^3 \delta_{\epsilon_i,-1} \bfe_i.
\end{eqnarray*}
The symmetry group of the Cl atom is $H=\{\cgop{R_p}{{\bftm}_p}, R_p\in G_0\}$
with one ${\bftm}_p$ in
$T_H=\{0,\bfe_x,\bfe_y,\bfe_z,
\bfe_x+\bfe_y,\bfe_y+\bfe_z,\bfe_x+\bfe_z,
\bfe_x+\bfe_y+\bfe_z\}$ for each $R_p$.
The groups $K$ and $H$ are both isomorphic to $G_0$ although they 
contain different elements of $G$.

\subsubsection{Double cosets}
\label{halitedoublecosetsect}
To determine the double cosets 
$H\backslash G/K$ we observe that, if $g_j=\cgop{\id}{\bft_j}$ with $\bft_j$ in $\Lambda$,
\begin{eqnarray*}
g_p g_j g_q &=& \cgop{R_pR_q}{R_p\bft_j+{\bftm}_p}.
\end{eqnarray*}
To compute the double coset $H g_j K$ we take into account that
$R_q$ is arbitrary, so that $R_pR_q$ takes all values in $G_0$
for any value of $p$. The number of elements of 
$H g_j K$
is 48 times the number of different values of 
${R_p\bft_j+{\bftm}_p}$ when $\cgop{R_p}{{\bftm}_p}$
runs over $H$, which varies from 6 (for $\bftn=\bfa_i$) to 48 
according to the specific form of $\bftn$.
We recover the fact that, contrary to cosets,
double cosets can have different sizes.

Let us choose a representative of the form $g_j=\cgop{\id}{\bft_j}$ of 
$H\backslash g_j/K$.
In the reference frame of the conventional cell,
if $\bft_j=(r_1,r_2,r_3)$ the 
elements of $H\backslash g_j/K$
are $g=\cgop{R}{\bftm}$ such that
$\bftm = {R_p\bft_j+{\bftm}_p}$, which
are the vectors whose coordinates are
all permutations of each triple of the following set:
$\{(r_1,r_2,r_3)$, $(1-r_1,r_2,r_3)$, $(r_1,1-r_2,r_3)$, 
$(r_1,r_2,1-r_3)$, $(r_1,1-r_2,1-r_3)$, $(1-r_1,r_2,1-r_3)$, 
$(1-r_1,1-r_2,r_3)$, $(1-r_1,1-r_2,1-r_3)\}$.
One and only one of these vectors
has coordinates $(t_1,t_2,t_3)$ such that
$0 < t_1\le t_2\le t_3$. 
Indeed, we can choose all coordinates
to be strictly positive because 
if any $r_i\le 0$ we can replace it by $1-r_i>0$
and then we can order these coordinates in increasing order
because all permutations are present.

With this particular
$\bft_j=(t_1,t_2,t_3)$, the operation $\cgop{\id}{\bft_j}$
is chosen to be the representative of $H\backslash g_j/K$.
Finally, the crystallographic group $G$ is the
union of $H\backslash g_j/K$ where $g_j=\cgop{\id}{\bft_j}$
runs over all $\bft_j=(t_1,t_2,t_3)$ such that
$0 < t_1\le t_2\le t_3$ and 
$\{t_1,t_2,t_3\}$ is either a set of three integers
or a set containing one integer and two half-odd-integers.

\subsubsection{Determination of $K_j$}

The next step of the restriction formula is to build the subgroups
$K_{j}=g_{j} K g_{j}^{-1} \cap H$ for all 
representatives $g_j=\cgop{\id}{\bft_j}$ of $H\backslash G/K$.
An element $\cgop{R_p}{{\bftm}_p}$ is in $K_j$
if there is a $g_q$ in $K$ such that $g_{j} g_q g_{j}^{-1}$ is in $H$
and
\begin{eqnarray*}
g_{j} g_q g_{j}^{-1} &=& \cgop{R_q}{-R_q\bft_j+\bft_j}
=\cgop{R_p}{{\bftm}_p}.
\end{eqnarray*}
Therefore, $R_q=R_p$ and 
$R_p\bft_j+{\bftm}_p=\bft_j$.
The first equality is not a restriction because both $K_0=H_0=G_0$.
Because of the discussion leading to the choice of $\bft_j$,
the symmetry groups corresponding to the different $\bft_j$ are:
\begin{enumerate}
\item $\bft_j=(1/2,1/2,n_3)$, $n_3>0$, $K_j=\{\cgop{(x,y,z)}{0}, \cgop{(-y,x,z)}{\bfe_x}, \cgop{(-x,-y,z)}{\bfe_x+\bfe_y}, \cgop{(y,-x,z)}{\bfe_y},\\
\cgop{(-x,y,z)}{\bfe_x}, \cgop{(-y,-x,z)}{\bfe_x+\bfe_y}, \cgop{(x,-y,z)}{\bfe_y}, \cgop{(y,x,z)}{0}\}$
isomorphic to $C_{4v}$ (or $4mm$),
\item $\bft_j=(1/2,n_2+1/2,n_3)$, $n_2>0$, $n_3>n_2$, 
$K_j=\{(\cgop{(x,y,z)}{0}, \cgop{(-x,y,z)}{\bfe_x}\}$, isomorphic to $C_s$ (or $m$),
\item $\bft_j=(1/2,n_2,n_3+1/2)$,  $n_2 > 0$, $n_3\ge n_2$,
$K_j=\{\cgop{(x,y,z)}{0}, \cgop{(-x,y,z)}{\bfe_x}\}$, isomorphic to $C_s$ (or $m$),
\item $\bft_j=(n_1,n_1,n_1)$,  $n_1 > 0$, 
$K_j=\{\cgop{(x,y,z)}{0}, \cgop{(z,x,y)}{0}, \cgop{(y,z,x)}{0}, 
\cgop{(y,x,z)}{0}, \cgop{(x,z,y)}{0}, \cgop{(z,y,x)}{0}\}$, isomorphic to
$C_{3v}$ (or $3m$)
\item $\bft_j=(r_1,r_1,n_3)$,  $r_1 \ge 1$, $n_3>r_1$,
$\{K_j=\cgop{(x,y,z)}{0}, \cgop{(y,x,z)}{0}\}$, isomorphic to $C_s$ (or $m$),
\item $\bft_j=(n_1,r_2,r_2)$,  $n_1 \ge 1$, $r_2>n_1$, 
$K_j=\{\cgop{(x,y,z)}{0}, \cgop{(x,z,y)}{0}\}$, isomorphic to $C_s$ (or $m$),
\item $\bft_j=(r_1,r_2,r_3)$,  $1\le r_1<r_2<r_3$, 
$K_j=\{(\cgop{(x,y,z)}{0}\}$, isomorphic to $C_1$ (or $1$),
\end{enumerate}
where $n_i$ are integers and $r_i$ half integers (i.e. integers or half-odd-integers).

\subsubsection{Restriction formula}

For an irreducible representation $\rho$ of $K$
and a representative $g_j=\cgop{\id}{\bft_j}$ of $H\backslash G/K$,
recall that a representation of $K_j$ is 
$\rho_j(k)=\rho(g_j^{-1}kg_j)$ with $k\in K_j$.
Recall also that an element of $K_j$ is an element 
$k=\cgop{R_p}{{\bftm}_p}$ of $H$ such that 
$R_p\bft_j+{\bftm}_p=\bft_j$
Therefore, 
\begin{eqnarray*}
g_{j}^{-1} k g_{j} &=& \cgop{\id}{-\bft_j}\cdot\cgop{R_p}{{\bftm}_p}\cdot\cgop{\id}{\bft_j}
=\cgop{\id}{-\bft_j}\cdot\cgop{R_p}{R_p\bft_j+{\bftm}_p}
=\cgop{\id}{-\bft_j}\cdot\cgop{R_p}{\bft_j}
=\cgop{R_p}{\bft_j-\bft_j}
\\&=&\cgop{R_p}{0},
\end{eqnarray*}
which is indeed an element of $K$. The result is simple, $g_{j}^{-1} k g_{j} $ is just 
the rotation part of $k$.

The electronic structure of Na is $1s^2 2s^22p^63s^1$ 
and the electronic structure of Cl is 
$1s^2 2s^22p^63s^23p^5$. 
The question is how the $3s$ valence electron of Na can induce a representation of $G$
which restricts to the $3p$ hole of the valence shell of Cl. 
The $3s$ electron generates a fully symmetric $A_{1g}$ irreducible representation
of the $O_h$ symmetry group of Na,
while the $3p$ hole generates a $T_{1u}$ irreducible representation of
$O_h$ symmetry group of Cl.

The relevant irreducible representation of $Na$ is the one-dimensional fully symmetric irrep
$(V,\rho)$ for which $\rho(g)=1$ for all $g\in K$. 
For every $K_j$, the representation $\rho_j(k)=\rho(g_j^{-1}kg_j)$ is again
the one-dimensional fully symmetric irrep of $K_j$. 
Therefore, the representation of $H$ induced from $\rho_j$ is the regular representation
of $H$. 
By definition~\eqref{inducedrep}, the induced representation $\ind_{K_j}^H\rho_j$ of the
trivial representation of $K_j$ is the permutation representation
of the cosets of that subgroup. 
\begin{eqnarray*}
\big(\tau_j(h)\big)_{m,n} &=& 
\sum_{k\in K_j} \delta_{k,h_m h h_n^{-1}},
\end{eqnarray*}
where $h_m$ and $h_n$ run over a full set of representatives in $H$ of the (left) cosets of $K_j$.
Its character is 
\begin{eqnarray*}
\chi_j(h) &=& 
\sum_m\sum_{k\in K_j} \delta_{k,h_m h h_m^{-1}}.
\end{eqnarray*}
It just remains to determine the number of times $T_{1u}$ appears 
in the induced representation $\ind_{K_j}^H\rho_j$ for each $K_j$.
The decomposition of these induced representations 
into irreps of $O_h$ for the different $K_j$ is
(in the same order as the enumeration of $\bft_j$ above)
\begin{enumerate}
\item $K_j = C_{4v}$ : $A_{1g} \oplus E_g \oplus T_{1u}$,
\item $K_j = C_s$ : $A_{1g} \oplus A_{2g} \oplus 2\times E_g \oplus T_{1g} 
\oplus T_{2g} \oplus 2\times T_{1u} \oplus 2\times T_{2u}$,
\item $K_j = C_s$ : $A_{1g} \oplus A_{2g} \oplus 2\times E_g \oplus T_{1g} 
\oplus T_{2g} \oplus 2\times T_{1u} \oplus 2\times T_{2u}$,
\item $K_j = C_{3v}$ : $A_{1g} \oplus T_{2g} \oplus A_{2u} \oplus T_{1u}$,
\item $K_j = C_s$ : $A_{1g} \oplus E_g \oplus T_{1g} \oplus 2\times T_{2g} 
\oplus A_{2u} \oplus E_u \oplus 2\times T_{1u} \oplus T_{2u}$,
\item $K_j = C_s$ : $A_{1g} \oplus E_g \oplus T_{1g} \oplus 2\times T_{2g} 
\oplus A_{2u} \oplus E_u \oplus 2\times T_{1u} \oplus T_{2u}$,
\item $K_j = C_1$ : $A_{1g} \oplus A_{2g} \oplus 2\times E_g \oplus 3\times T_{1g} 
\oplus 3\times T_{2g} \oplus A_{1u} \oplus A_{2u} \oplus 2\times E_u  
\oplus 3\times T_{1u} \oplus 3\times T_{2u}$.
\end{enumerate}
Note that subgroups $C_s$ can give different restrictions, depending
on the group operations used to build them~\cite[p.82]{Altmann}.

Here we are interested in the coupling of the representation induced
by the 3s orbital of Na with the representation $T_{1u}$ of the
3p orbital of Cl. Since the Hamiltonian of the system is invariant
under the action of the space group, the coupling with
the $T_{1u}$ irrep of Cl of Na is only possible with 
the  $T_{1u}$ irrep of the restriction of the representation induced
from the Na orbital. Mackey's restriction formula gives us
a complete answer to the question of the number of such couplings, which is
1, 2, 2, 1, 2, 2, 3 for the different $K_j$. This is also the number
of parameters used to describe the coupling in a phenomenological approach.
Of course, the intensity of the coupling depends on the distance
between the Na and Cl atoms.


\subsection{Mackey's decomposition formulas}
For the investigation of selection rules and Clebsch-Gordan coefficients,
it is necessary to have a precise description of the tensor 
product of two irreducible representations. We observed that 
irreducible representations of crystallographic groups can always
be induced from irreducible representations of subgroups. 
Therefore, it is useful to describe the tensor product
of induced representations. 

As a preparation for Mackey's decomposition formula,
we start with a simpler situation. If $H$ is a closed subgroup of a locally compact group $G$,
then for arbitrary representations $\rho$ of $G$ and $\tau$ of $H$
the following relation holds~\cite[Thm.~2.58]{Kaniuth-13}
\begin{eqnarray*}
\rho\otimes\ind_H^G \tau &=& \ind_H^G(\rho|_H \otimes \tau).
\end{eqnarray*}
We want to determine what happens when $\rho$ itself is an
induced representation, in other words, we want to decompose
the tensor product of two induced representations
in terms of induced representations.

As we shall see, this problem was solved by \emph{Mackey's decomposition formula}
for finite groups~\cite{Mackey-51} and for some infinite groups~\cite{Mackey-52}, under the
same conditions as for Mackey's restriction formula. Therefore,
this formula holds for crystallographic groups. 
Mackey's decomposition formula was used (without proof)
by Bradley~\cite{Bradley-66} to determine
selection rules and by Patricia Gard~\cite{Gard-73} 
and Rhoda Berenson~\cite{Berenson-75}
to compute Clebsch-Gordan coefficients of space-group representations. Since then,
Mackey's method is commonly used because of the simplification 
it brings to the understanding and calculation of 
Clebsch-Gordan coefficients~\cite{Broek-79-II,Davies-86,Davies-86-II,Davies-86-III,Elcoro-20}.

Let $G$ be a crystallographic group and $H$ and $K$ two subgroups of $G$.
Let $g_j$, with $j\in J\subset\bbN$  (where $J$ can be infinite) 
be a complete set of representatives 
of the double cosets $H\backslash G/K$, so that 
$G=\cup_{j\in J} H g_j K$. Define again the sugroups $K_j=g_j K g_j^{-1} \cap H$
of $H$.
If $(\calH,\rho)$ is a representation of $K$, then
$\rho_j(k)=\rho(g_j^{-1}kg_j)$, for $k\in K_j$, is a representation of $K_j$.
If $\sigma$ is a representation of $H$,
then Mackey's decomposition formula states that~\cite[Thm.~7.2]{Mackey-52}
\begin{eqnarray*}
\ind_H^G\sigma\otimes \ind_K^G\rho &=& 
\bigoplus_{j\in J} \ind_{K_j}^G\big(\sigma|_{K_j}\otimes \rho_j\big).
\end{eqnarray*}
In words, the left hand side of this formula
is the tensor product of the representations of $G$ induced
from the representation $\sigma$ of $H$ and the
representation $\rho$ of $K$. 
For the right hand side, we consider the subgroup $K_j$ of 
$H$, we make the tensor product
of the restriction of $\sigma$ to $K_j$ by the
representation $\rho_j(k)=\rho(g_j^{-1}kg_j)$ of $K_j$,
we induce a representation of $G$ from this tensor
product, and we sum over $j$. 

For future reference, we write the case where $K=H$ and $\sigma=\rho$:
\begin{eqnarray*}
\ind_H^G\rho\otimes \ind_H^G\rho &=& 
\bigoplus_{j\in J} \ind_{H_j}^G\big(\rho|_{H_j}\otimes \rho_j\big),
\end{eqnarray*}
where $H_j=g_j H g_j^{-1} \cap H$ and 
$\rho_j(h)=\rho(g_j^{-1}hg_j)$ is a representation of  $H_j$.

Bradley~\cite{Bradley-66} made a detailed application of Mackey's decomposition formula
for the irreducible representations of space group $P23$
(N$^\circ$~195 in Ref.~\cite{Hahn}).
Wawrzycki~\cite{Wawrzycki-19} extended these formulas to the case of Krein spaces 
(a generalization of Hilbert space useful in quantum field theory~\cite{Bizi-18}).

\subsection{Mackey's decomposition of symmetric and antisymmetric squares}
Symmetric and antisymmetric squares of representations have
many applications in quantum physics. 
The most common ones come from the fact that antisymmetric squares describe
fermions (e.g. coupled electrons) while symmetric squares
describe bosons (e.g. coupled phonons).
For example, they were used 
to investigate Cooper pairs~\cite{Yarzhemsky-04,Yarzhemsky-25} 
or unconventional gaps~\cite{Sumita-18} 
in superconductors, the coexistence of magnetism and 
superconductivity~\cite{Nomoto-17},
and soft phonon modes in antiferromagnets~\cite{Cracknell-70}).

Bradley describes~\cite{Bradley-70} the use of symmetric and antisymmetric squares to compute 
transitions between an initial and a final states when the final state of
is related to the initial state by time
reversal (see also~\cite[Sect.~4.8]{Bradley-10}).
Finally, a group $G$ is said to be multiplicity-free when
the decomposition of the tensor product of two irreducible representations of $G$
does not contain any irreducile representation of $G$ more than once
(simply reducible groups listed by Luan~\cite{Luan-20} are
multiplicity-free).
Dealing with non multiplicity-free groups is possible~\cite{Derome-65,Butler} but
keeping track of this multiplicity generates a cluttering of indices.
Fortunately, $SO(3)$ and all three-dimensional crystallographic point groups
are multiplicity free, except for the tetrahedral groups $T$ and $T_h$
for which the square of the irrep $T$ (not to be confused with the group $T$)
contains twice the irrep $T$~\cite[p.~593]{Altmann}:
\begin{eqnarray*}
	T\otimes T &=& A \oplus {}^1E \oplus {}^1E \oplus T \oplus T.
\end{eqnarray*}
These two $T$ can be uniquely specified
by requiring that one is in the symmetric square of $T$
and the other in its antisymmetric square.
Bradley discussed the case of finite groups and we extend his
work to infinite crystallographic groups.

In the case of a \emph{finite} group $G,$ Mackey  proved a decomposition formula for
the symmetric and antisymmetric squares of an induced representation of $G$ 
as a sum of induced representations~\cite[Thm. 1]{Mackey-53} 
(see also~\cite[Thm.~12.13]{Curtis-I}).
The case of classical groups was treated by Magaard et al~\cite{Magaard-02}.
At the end of his paper, Mackey writes that he intends
to discuss the case of infinite groups in ``a subsequent paper" which, 
as far as we know, was never published.

We prove now that his formulas remain  valid for crystallographic
groups and, more generally, for discrete countable groups.
We prove this in detail to illustrate the kind of work that
	has to be done to go from a finite-group property to
	its infinite-group counterpart.
Let $G$ be a discrete countable group; let $H$ be a subgroup of $G$ and 
$\rho$ a representation of $H$ on a Hilbert space $\mathcal K$.
Set 
$$\pi:= \ind_H^G\rho.$$
Recall that 
the (interior) tensor product  $\pi\otimes \pi,$ which is a representation of $G,$ can be 
identified
with the restriction to the diagonal subgroup $\Delta_G$ of $G\times G$
of the exterior tensor product $\pi \times \pi$, which is a representation of 
$G\times G.$ Now, $\pi \times \pi$ is equivalent to the induced representation
$\ind_{H\times H}^{G\times G} (\rho\times \rho).$
More explicitly, the direct product $G\times G$ is the
set of $(x,y)$ with $x$ and $y$ in $G$,
equipped with the group product 
$(x,y)\cdot(x',y')=(xx',yy')$ and its diagonal
subgroup is the set $\Delta_G=(x,x)$ for $x$ in $G$.
The exterior tensor product is also called
outer Kronecker product~\cite[Section~4.4]{LudwigFalter}.

The Hilbert space of $\rho\times \rho$, which is the tensor product $\mathcal{K}\otimes \mathcal{K},$
can be identified~\cite[Ch.~12]{Aubin} with the space
$\mathrm{HS}\overline{\mathcal{K}},\mathcal{K})$ of Hilbert-Schmidt operators
${\rm HS}(\overline{\mathcal{K}},\mathcal{K})$ of Hilbert-Schmid toperators
$T: \overline{\mathcal{K}} \to \mathcal{K}$, where $\overline{\mathcal{K}}$
is the conjugate Hilbert space 
and we have
$$(\rho (h_1) \times \rho(h_2))(T)=\rho(h_1) T \rho(h_2)^* \quad\text{for all}\quad h_1, h_2\in H.$$

	Recall that $\overline{\mathcal{K}}$ is the Hilbert space which agrees with $\mathcal{K}$
	as additive group, with multiplication by a scalar $\lambda$ being the 
	multiplication
	with $\overline{\lambda}$, and with the scalar product of $v,w\in \overline{\mathcal{K}}$
	being $\overline{\langle v| w\rangle}.$
	Recall also that  the identification of $\mathcal{K}\otimes \mathcal{K},$
	with ${\rm HS}(\overline{\mathcal{K}} , \mathcal{K})$ is given by the linear
	extension of the map $v\otimes w\mapsto (u\mapsto \langle u| v\rangle w).$
	
So, the Hilbert space $\mathcal{H}$ of $\pi \otimes \pi$
can be realized as the space of measurable maps
$F: G\times G\to {\rm HS}(\overline{\mathcal{K}} , \mathcal{K})$ such that
$$F(h_1x, h_2y)= \rho(h_1)F(x,y)\rho(h_2)^* \quad\text{for all} \quad h_1, h_2\in H, x,y\in G,$$
and for which  the function $(x,y)\mapsto \Vert F(x,y)\Vert$
(which is constant on the right cosets of $H\times H$
in $G\times G$) is square integrable on $H\times H\backslash G\times G;$
we then have 
$$(\pi(g) \otimes \pi(g))(F)(x,y)= F(xg, yg)\quad\text{for all} \quad  g\in G.$$
We now describe the elements in $\mathcal{H}$ which correspond to the symmetric and antisymmetric
	tensors. The Hilbert  space of $\pi$ is  the space $\calH_\pi$ of measurable maps
	$\Phi: G\to\mathcal{K}$ such that 
$$\Phi(hx)= \rho(h)\Phi(x)\ \quad\text{for all} \quad h \in H, x\in G,$$
	and for which  the function $x\mapsto \Vert \Phi(x)\Vert$
	is square integrable on $H\backslash G.$ The identification
of $\calH_\pi \otimes \calH_\pi$ with $\mathcal{H}$ is given by the linear extension of the map
	$\Phi_1 \otimes \Phi_2\mapsto F,$ with 
$$F(x,y): u\mapsto \langle u| \Phi_1(x)\rangle \Phi_2(y)\quad\text{for all} \quad x,y \in G, u\in \overline{\mathcal{K}}.$$
	Since the adjoint of the map $u\mapsto \langle u| v\rangle w$ is 
	$u\mapsto \langle u|w\rangle v,$
the map associated to $\Phi_2 \otimes \Phi_1$ is  $\check{F}\in \mathcal{H}$ given by
	$\check{F}(x,y)= F(y,x)^*.$
	It follows that the space of symmetric, respectively antisymmetric, tensors 
corresponds to the space of all $F\in\mathcal{H}$ with
$F=\check{F}$, respectively with $F=-\check{F}$.

For a double-coset $d\in H\times H\backslash G\times G/ \Delta_G$, let 
$\mathcal{H}_d$ be the subspace of  maps $F\in \mathcal{H}$
with $F=0$ outside $d.$ 
Let $d\in  H\times H\backslash G/ \Delta_G$
and $(e, a^{-1})\in d.$ Set  
$\rho_a (h)= \rho(a^{-1} h a )$ and let 
$\sigma_a$
be the  restriction of  
$\rho \otimes \rho_a$
to $H_a=H\cap aH a ^{-1}$.
Set $\tau_a=\ind_{H_a}^G \sigma_a$.
Mackey's decomposition formula for $\pi\otimes \pi$ can be summarized as follows:
\begin{enumerate}
	\item  each $\mathcal{H}_d$ is invariant under $\pi \otimes \pi$;
	\item  $\mathcal{H}$ is the direct sum of the $\mathcal{H}_d$'s 
	where $d$ ranges  over $H\times H\backslash G/ \Delta_G;$
	\item  
	the restriction of $\pi \otimes \pi$ to $\mathcal{H}_d$ is equivalent 
	to  $\tau_a$ for any $a\in G$ such that $(e, a^{-1})\in d$ (for an explicit equivalence
	between these representations,  see  below).
\end{enumerate}
We will identify the double-coset space 
$H\times H\backslash G\times G/ \Delta_G$ with the double-coset space $H\backslash G/ H$
by means of  the bijection
$$
(H\times H)(x,y) \Delta_G \mapsto H xy^{-1} H.
$$

In the above model for $\mathcal{H},$ 
the representation $\pi \otimes_s \pi$
respectively $\pi \otimes_a \pi$
coincides with the restriction of $\pi\otimes \pi$
to  the invariant subspace of all $F\in\mathcal{H}$ with
$F(x,y) =F(y,x)^*$ respectively with $F(x,y) =-F(y,x)^*$).
So, for every double-coset $d=Ha^{-1} H$,  the space $\mathcal{H}_d+ \mathcal{H}_{d^{-1}}$ is 
invariant under $\pi \otimes_s \pi$
as well as under  $\pi \otimes_a \pi$, where 
$d^{-1}=Ha^{-1}H.$
Of course, we have 
$$ \pi\otimes \pi= (\pi \otimes_s \pi) \oplus (\pi \otimes_a \pi).$$

We choose representatives for the double-coset space $H\backslash G/H$
and divide them into three classes as follows: we choose the unit $e$ of $G$ as representative of the class 
$H$; let $\{a_i\}_{i\in I}$ be a complete set of the representatives
of $H\backslash G/H$, different from $H$ and
such that $a_i^{-1}$  belongs to the same double-coset as $a_i$;
let $\{b_j\}_{j\in J}$ be a complete set of the representatives of 
$H\backslash G/H$ for which $b_j$ and $b_j^{-1}$ do not belong to the same class.

For every $j$ in $J$,
set 
$$\mathcal{H}_j:=\mathcal{H}_{d_j}\oplus \mathcal{H}_{d_j^{-1}}$$
for $d_j= Hb_jH.$
Then  $\mathcal{H}_j$ decomposes as the direct sum of two subspaces,
one invariant under $\pi \otimes_s \pi$ and the other under 
$\pi \otimes_a \pi;$ it turns out that the representations
of $G$ defined by these subspaces are both equivalent to 
of $\tau_j:=\ind_{H_{b_j}}^G \sigma_{b_j}$. Indeed, the proof is exactly the same as the
the one for Case I in ~\cite[p.391-392 ]{Mackey-53}.

For the double-coset $d_e=H$ corresponding to $e,$
the representation of $G$ on  $\mathcal{H}_{d_e}$
is the direct sum of a copy of $\pi \otimes_s \pi$ and a copy of $\pi \otimes_a \pi;$
this corresponds to  Case II in ~\cite[p.393-394 ]{Mackey-53}.

The third case case is more complex  (this corresponds to Case III in ~\cite[p.394 ]{Mackey-53}).
Let $i\in I$.
Since $a_i$ is a representative of 
a self-inverse double-coset,
$Ha_i H=H a_i^{-1}H$ and there exist
elements $h_1,\dots, h_4$ of $H$ such that
$h_1 a_i h_2=h_3 a_i^{-1} h_4$.
As a consequence, $a_i h_2h_4^{-1}=h_1^{-1}h_3 a_i^{-1}$
and the set 
$a_i H\cap H a_i^{-1}$ is not empty.
Let $c$ be an element of this set:
$$c=a_i h=h'a_i^{-1}$$
for some $h$ and $h'$ in $H$.
It is clear that $c$ is not in $H$
(because $a_i$ is not in the double-coset $H$)
and $c^2$ is in $H_{a_i} =H\cap a_i H a_i^{-1}$.
Moreover $cH_{a_i} c^{-1}= H_{a_i},$ since 
$c H c^{-1}=a_i hHh^{-1}a_i^{-1}= a_i H a_i^{-1}$
and $c (a_i H a_i^{-1})c^{-1}=h'a_i^{-1}(a_i H a_i^{-1})a_i(h')^{-1}= H.$
As a consequence, $K:=H_{a_i} \cup c H_{a_i}$ is a subgroup of $G$
that contains $H_{a_i}$ as an invariant subgroup of index 2.

We first express in terms of $c$ the   involution
defining the symmetric and 
antisymmetric tensors, restricted to $\mathcal{H}_{d_{a_i}}.$
Recall that  $\sigma_{a_i}$  is the  restriction of  
$\rho \otimes \rho_{a_i}$
to $H_{a_i}.$
The Hilbert space of $\sigma_{a_i}$ is the space ${\rm HS}(\overline{\mathcal{K}}, \mathcal{K})$  of Hilbert Schmidt operators 
from $\overline{\mathcal{K}}$ to  $\mathcal{K}.$ 
Let $\mathcal{H}_0$ be the  Hilbert space  of measurable maps
$F_0:  G\to {\rm HS}(\overline{\mathcal{K}}, \mathcal{K})$ such that 
$$F_0(h x)= \rho(h)F_0(x) \rho (a_i^{-1} h a_i)^* \quad\text{for all}\quad h\in H_{a_i}, x\in G,$$ 
and for which the function $x\mapsto \Vert F_0(x)\Vert$
is square integrable on $H_{a_i}\backslash G.$ The group $G$ acts
on $\mathcal{H}_0$ by right translation; in fact, 
$\mathcal{H}_0$  is a realization of the Hilbert space of 
$\tau_{a_i}=\ind_{H_{a_i}}^G \sigma_{a_i}$.
To  every $F_0\in \mathcal{H}_0,$ we can associate $F\in \mathcal{H}_{d_{a_i}}$
defined by 
$$
F(h_1x,h_2 a_i^{-1} x)= \rho(h_1)F_0(x) \rho(h_2)^*  \quad\text{for all}\quad h_1,h_2\in H, x\in G;
$$
the map $F_0\mapsto F$ is an isomorphism between $ \mathcal{H}_0$
and  $\mathcal{H}_{d_{a_i}}$ intertwining the representations $\tau_{a_i}$
and  $\pi\otimes \pi$.
The involution  $F\mapsto \check{F}$ 
on $\mathcal{H}_{d_{a_i}}$ corresponds to the 
involution 
${\rm Inv}_0$
on $ \mathcal{H}_0$ given by 
$$
{\rm Inv} (F_0)(x)=  \rho(c^{-1} a_i) F_0(cx)^*\rho( a_i^{-1}c^{-1})^* \quad\text{for all}\quad  x\in G
$$
(see \cite[Lemma a ]{Mackey-53}).

Let 
$$\Phi: {\rm HS}(\overline{\mathcal{K}}, \mathcal{K})\to {\rm HS}(\overline{\mathcal{K}}, \mathcal{K})$$ be  defined by 
$$
\Phi(S)= \rho(c^{-1}a_i)S^*  \rho( a_i^{-1} c^{-1})^*
= \rho(c^{-1} a_i )S^*  \rho(  ca_i ) \quad\text{for all}\quad S\in {\rm HS}(\overline{\mathcal{K}}, \mathcal{K}).
$$
(Observe that 
$\rho( c^{-1}a_i)$ and $\rho(a_i^{-1} c^{-1})$ are well defined since
$c^{-1}a_i$ and $a_i^{-1} c^{-1}$ are elements of  $H$.).
Set 
$$\sigma_i^{+}(c)=\Phi^{-1}\quad \text{ and}\quad  \sigma_i^{-}(c)=-\Phi^{-1};$$
and $\sigma_i^{\pm} (h)= \sigma_{a_i} (h)$ for $h\in H_{a_i}.$
Then $\sigma_{a_i}^{+}$ and $\sigma_{a_i}^{-}$  are two representations
of $K$
on ${\rm HS}(\overline{\mathcal{K}}, \mathcal{K})$ extending $\sigma_i.$
Indeed, one checks  that 
$$
(\pm \Phi^{-1})\sigma_{a_i}(h) (\pm \Phi^{-1})= \sigma_{a_i} (chc^{-1})  \quad \text{for all }\quad h\in H_{a_i}
$$
and that $(\pm \Phi^{-1})^2=\sigma_{a_i}(c^2)$ since
\begin{align*}
	(\pm \Phi^{-1})^2(S)&= \Phi^{-1}( \rho(c a_i )S^*  \rho(  c^{-1}a_i ) )\\
	&= \rho(ca_i)( \rho(( \rho(c a_i )S^*  \rho(  c^{-1}a_i ))^*(\rho(c^{-1}a_i)\\
	&=\rho(c^2 )S \rho (a_i^{-1} c^{-2} a_i)\\
	&=\rho(c^2 )S \rho (a_i^{-1} c^{2} a_i)^*\\
	&=\sigma_{a_i}(c^2)(S).
\end{align*}
Let $F_0\in \mathcal{H}_0.$ Observe that
${\rm Inv} (F_0)(x)= \Phi(F_0(cx))$ and hence
${\rm Inv} (F_0)=F_0$ if and only if 
$$F_0(cx)= \Phi^{-1}(F_0(x))= \sigma_i^{+}(c) (F_0(x))  \quad \text{for all }\quad x\in G. $$
Since $F_0(hx)= \sigma_{a_i}(h) (F_0(x))$ for all $h\in H_{a_i},$ if follows
that ${\rm Inv} (F_0)=F_0$ if and only if 
$$
F_0( kx)= \sigma_{a_i}^+(k) (F_0(x)) \quad \text{for all }\quad k\in K, x\in G
$$
This shows that the subrepresentation of $\pi \otimes \pi$ (realized on  $\mathcal{H}_0$)  corresponding to 
$\pi \otimes_s \pi$ 
is equivalent to the induced representation $\tau_i^{+}:= \ind_K^G \sigma_{a_i}^{+}.$
Similarly, the subrepresentation of $\pi \otimes \pi$   corresponding to 
$\pi \otimes_a \pi$ 
is equivalent to the induced representation $\tau_i^{-}:= \ind_K^G \sigma_{a_i}^{-}.$

As final result, one obtains the following decompositions

\begin{eqnarray*}
	(\ind_H^G\rho) \otimes_s (\ind_H^G\rho) &=& 
	\ind_H^G(\rho\otimes_s\rho) \oplus \bigoplus_{i\in I}\tau_i^+ \oplus \bigoplus_{j\in J} \tau_j,\\
	(\ind_H^G\rho) \otimes_a (\ind_H^G\rho) &=& 
	\ind_H^G(\rho\otimes_a\rho) \oplus \bigoplus_{i\in I} \tau_i^- \oplus \bigoplus_{j\in J} \tau_j.
\end{eqnarray*}

Bradley~\cite{Bradley-70} gives a detailed exemple of the use of these formulas
for the case of sphalerite (ZnS, the most common ore of zinc, 
a crystal with space group $F\bar{4}3m$, No~216). 
Detailed computations of symmetrized squares of 
space group irreps were carried out~\cite{Birman-62,Chen-65}.
Further properties of induced representations are described
in a readable way by Ceccherini-Silberstein et al.~\cite{Ceccherini-09,Ceccherini-15}.

Note that Mackey's decomposition of tensor squares of induced representations was 
generalized to the permutation symmetry decomposition of any tensor power of 
an induced representation in a remarkable paper by Patricia 
Gard~\cite{Gard-73-induced-power} and then applied to space group representations~\cite{Gard-73-space-group-power}.

\section{Conclusion}
This paper is a contribution to the formulation of the physics of
infinite crystals, avoiding periodic boundary conditions and questions
about the validity of the thermodynamic limit.
Our purpose was to close the gap between the description of the 
irreducible representations of the symmetry group of (finite) toric crystals
and the use of infinite crystals required
to investigate the analytical properties of eigenstates and eigenenergies
of band structure calculations. 
We presented here the properties of irreducible representations
of infinite crystals that are similar to those
 of (finite) toric crystals.

The present work can be extended in several directions: by investigating
more general groups or by studying other properties of group representations.
For example, spin-space groups were first considered by Brinkman 
and Elliott~\cite{Brinkman-66} and they 
play an important role in the study of electronic band structure~\cite{Sandratskii-79}
of magnon band topology~\cite{Corticelli-22} and altermagnetism. 
Spin-space groups (in dimension 3) were recently enumerated~\cite{Chen-24,Jiang-24,Xiao-24}.
By generalizing the case of 3-dimensional spin-space groups~\cite{Liu-22}
we can define  spin-space groups in dimension $n$ as any subgroup of the direct product
$G\times SO(n)\times \bbZ_2$, where $G$ is a crystallographic group
and $SO(n)\times \bbZ_2$ acts on the spin-variable. Since $SO(n)$ is a 
compact group,  the derivation of the irreducible representations of
spin-space groups for infinite crystals does not
meet additional difficulty. Note that $SO(n)\times \bbZ_2$
can be replaced by any compact group, for instance $U(n)$.
Our results on magnetic space groups can also be
extended to colored space groups~\cite{Lifshitz}
and the related Mackay groups~\cite{Mackay-57,Jablan-02}.
It is possible to define induced corepresentations of magnetic 
groups~\cite{Evarestov-97,Rumynin-21}. It would be interesting to
extend Mackey's restriction and decomposition formula to that case. 

The results of the present paper may give the impression that 
infinite crystals are very much like finite toric crystals
and that the presence of an infinite number of elements
and of irrational-valued coordinates in the Brillouin
zone is harmless. 
This is however not the case for objects such as the characters of these representations,
which are no longer periodic functions but become ergodic
as in the example of $f(n)=e^{2\pi i\alpha n}$ in the introduction. 
Since character theory is such a powerful tool for finite groups,
it would be interesting to investigate how much of it can
be adapted to infinite crystallographic groups.
We intend to discuss this question in a forthcoming publication.

\section{Acknowledgements}
It is a pleasure to thank Coraline Letouz\'e for a discussion that lead to the present work, 
Bernd Souvignier for his help with Table~\ref{table-groupe},
Delphine Cabaret for sharing her vast knowledge of space group theory
and Vladimir Dmitrienko and Slava Chizhikov for exclusive information
about magnetic space groups.

\section*{References}

\bibliographystyle{plain}

\bibliography{qed}

\end{document}